\documentclass[12pt,a4paper]{article}
\usepackage{amsmath,amssymb,mathrsfs,framed,esint,slashed,braket}
\usepackage[colorlinks]{hyperref}
\usepackage{color}
\usepackage[all]{xy}
\usepackage{graphicx}
\usepackage[color=yellow]{todonotes}

\newcommand{\rem}[1]{}

\newcommand{\de}{{\rm d}}

\newcommand{\bq}{{\boldsymbol{q}}}

\newcommand{\br}{{\boldsymbol{r}}}

\newcommand{\bu}{{\boldsymbol{u}}}

\newcommand{\beq}{\begin{equation}}
\newcommand{\eeq}{\end{equation}}
\newcommand{\ben}{\begin{eqnarray}}
\newcommand{\een}{\end{eqnarray}}

\begin{document}

\title{
\vspace{-.25cm}The bohmion method in\\nonadiabatic quantum hydrodynamics
}
\author{Darryl D. Holm$^1$, Jonathan I. Rawlinson$^2$, Cesare Tronci$^{3,4}$
\smallskip
\\ 
\small
$^1$\it Department of Mathematics, Imperial College London, London, UK
\\ 
\small
$^2$\it School of Mathematics, University of Bristol, Bristol, UK
\\
\small
$^3$\it Department of Mathematics, University of Surrey, Guildford, UK
\\
\small
$^4$\it Department of Physics and Engineering Physics, Tulane University, New Orleans LA, USA}
\date{}
\maketitle

\begin{abstract} 
Starting with the exact factorization of the molecular wavefunction, this paper presents the results from the numerical implementation in nonadiabatic molecular dynamics of the recently proposed bohmion method. {Within the context of quantum hydrodynamics,} we introduce a regularized nuclear Bohm potential  admitting solutions comprising a train of $\delta-$functions  which provide a finite-dimensional sampling of the hydrodynamic flow paths. The bohmion method inherits all the basic conservation laws from its underlying variational structure and captures electronic decoherence. After reviewing the general theory, the  method is applied to the well-known Tully models, {which are used here as benchmark problems. In the present case of study, we show that the new method  accurately reproduces both electronic decoherence and nuclear population dynamics}. 
\end{abstract}


{
\tableofcontents
}

\bigskip

\section{Introduction\label{sec:intro}}

Among the {various} pictures of quantum mechanics, Madelung's hydrodynamics offers the invaluable advantage of preserving the concept of trajectory, which is lost in other pictures. Madelung's picture of quantum hydrodynamics (QHD) has been attracting much attention over the decades. Already serving as the bridge between the nonlinear Schr\"odinger equation and the dynamics of quantum fluids, QHD has recently been finding new applications ranging from quantum plasmas \cite{MoBoRa18} to the description of supersolid crystals \cite{HeBuDu19}.

The quantum potential arising in Madelung's transformation from the linear Schr\"odinger equation to QHD produces interference effects which are seldom encountered in classical hydrodynamic models. These interference effects comprise the essence of de Broglie's pilot-wave perspective. (See the bouncing droplet experiments in \cite{Bush} for a pilot-wave perspective of classical hydrodynamics.) Thus, while restoring the concept of Lagrangian fluid parcel trajectory, QHD also carries wave mechanics features that can transcend classical fluid motion. In particular, the presence of gradients of the density in the quantum potential eliminates the standard single-particle trajectories which are seen in the classical case. 

Despite the computational difficulties associated with interference effects in the quantum potential, QHD still attracts much attention for the development of convenient reduced models. In chemical physics, several efforts continue to be addressed towards the use of quantum hydrodynamic trajectories in molecular dynamics beyond the mean-field model \cite{Prezhdo01,Tavernelli13}. Since the appearance of the surface-hopping method in the early 70's \cite{Tully71}, the increasing availability of computational power has fostered a series of different approaches for the simulation of nonadiabatic systems in quantum molecular dynamics. 
In this context, the nuclear response to the quantum electronic transitions poses major challenges, since  the mean-field approximation is generally unable to capture such effects accurately. In addition, both the mean-field model and the surface-hopping method  falter in describing electronic decoherence, even though several corrections have been proposed over the years \cite{Prezhdo99,Granucci10,Subotnik16}. {All these difficulties in capturing the various features of vibronic interactions are related to the long-standing problem of quantum-classical coupling \cite{AgCi07,BoGBTr19,GBTr20,Kapral,Sudarshan}.}

In molecular dynamics, the QHD framework remains attractive despite the availability of several conventional methods resorting to basis set expansions which exploit the standard properties of frozen Gaussian wavepackets \cite{Ben-Nun98,HuHe88,Makhov14,Shalashilin09}. However, as pointed out in \cite{JoIz18}, ``the existent formalisms using classical frozen-width Gaussian motion do not conserve the total energy''.  In some cases, the conservation of total probability also falters, and likewise for the momentum balance \cite{HaWeTrBNMa01}. Notice that these are not issues appearing at the level of the numerical discretization. Instead, these are limitations in the computational model itself \cite{Ha12} and any improvement in this regard requires increasing the dimension of the basis set. While basis set expansions have a long standing tradition, trajectory-based approaches  offer the opportunity of developing new models whose equations of motion can be designed to preserve the correct conservation laws. Even so, the emergence of highly irregular profiles of the quantum potential can produce nodal points, thereby making this avenue particularly challenging \cite{ZhMa03}. At present, this challenge has not yet been met successfully.  Following Wyatt's extensive work \cite{Wyatt}, Garashchuk and collaborators have designed several methods for approximating the quantum potential in different test cases \cite{Garashchuk10}. Similarly, Gu and Franco adopted quantum trajectories for describing system-bath interaction \cite{GuFr17}, while Curchod and Tavernelli \cite{EntropyReviewNonadiabatic,Curchod11} proposed blending QHD methods with the usual Born-Huang expansion \cite{BornHuang27} in nonadiabatic dynamics. 

Most recently, Gross and collaborators \cite{CoupledvsIndependent} achieved a new breakthrough by combining hydrodynamic trajectories with the \emph{exact factorization} formalism \cite{AbediEtAl2012,Gross2010}. Instead of focusing on the Born-Huang series expansion, this picture involves an alternative representation of the molecular wavefunction, expressed as follows:
\beq
\Psi(\boldsymbol{r},\boldsymbol{x},t)=\chi(\boldsymbol{r},t)\phi(\boldsymbol{x},t;\br)
\qquad\text{with}\qquad
\int |\phi(\boldsymbol{x},t;\br)|^2\,\de^3 x=1
\,.
\label{ExFact}
\eeq
The electronic function $\phi(\boldsymbol{x},t;\br)$ is taken to be square-integrable in the electronic coordinates $\boldsymbol{x}$ while it is parameterized by the nuclear coordinate $\br$. Although this representation is reminiscent of the adiabatic Born-Oppenheimer (BO) theory, here the electronic function depends explicitly on time.   The representation of the wavefunction in \eqref{ExFact} was first considered several decades ago  \cite{Hunter} (see also Section 11.1 in \cite{BBirula}), although its advantages in nonadiabatic dynamics had apparently not been recognized until much more recently.

As the current computational schemes based on quantum trajectories are still under development, questions about conservation laws have been considered only seldom \cite{GaRo04,GaRo19}. While  mean-field dynamics conserves all constants of motion by construction, 
 satisfactory results beyond mean-field models are currently achieved only for certain parameter ranges. In this paper, we develop a recent trajectory-based approach previously proposed by our team in the context of nonadiabatic quantum hydrodynamics \cite{FoHoTr19}. In particular, we present the benchmark implementation of a new closure scheme obtained by combining regularization methods and hydrodynamic variational principles. The new scheme has the following characteristics.
\begin{enumerate}
\item It is based on hydrodynamic quantum trajectories; 
\item It retains  basic conservation laws for all parameter ranges;
\item It accurately reproduces electronic decoherence effects.
\end{enumerate}
The variational structure of the new scheme is obtained by exploiting recent progress based on the exact factorization ansatz \eqref{ExFact}, which enables the nuclear and electronic wavefunctions to be treated differently in the nonadiabatic context. The scheme's conservation laws follow from the variational principle underlying the exact factorization. In addition, a convolution regularization of the density is adopted to mitigate the difficulties arising from the quantum potential and admit point-particle histories -- called \emph{bohmions} -- which approximate the hydrodynamic paths. The regularization of the density avoids the singular limit $\hbar\to0$ which emerges in standard quantum hydrodynamics. Indeed, in the regularized dynamics, this limit is no longer singular and it can be treated on an equal footing with the case $\hbar\neq0$. In the latter case, non-zero $\hbar$ acts as a coupling constant for interactions among the (bohmion) particle histories. These interactions are  then responsible for nuclear wave-packet splitting and electronic decoherence.
Moreover, based on standard techniques in geometric mechanics, the bohmion closure introduces a singular momentum map \cite{HoMa04,HoTr2009} which is dual to the standard Madelung transformation of quantum mechanics \cite{Madelung1926,Madelung1927} into the language of hydrodynamics.  First formulated in \cite{FoHoTr19}, this geometric mechanics approach reveals the Lagrangian-particle content of quantum hydrodynamics. Namely, the singular bohmions follow Lagrangian flow trajectories in the regularized quantum hydrodynamics.

The content of the paper is as follows. In  Section \ref{sec:Bohmiontheory} we summarise the main points of the derivation of the bohmion model. By using the  exact factorization representation,  we set up the variational principle underlying nonadiabatic quantum hydrodynamics before applying the bohmion method to yield the bohmion equations of motion. In Section \ref{BohmionMeth} the bohmion method is derived along with its equations of motion. These equations are the starting point for the numerical simulations which follow. Section \ref{sec:results} contains new results from the numerical implementation of the bohmion method to the celebrated Tully models \cite{Tully}, with a focus on population transfer and decoherence dynamics. {Many of} these numerical implementations display good agreement with exact quantum mechanical results. In particular, they show that bohmions are able to capture electronic decoherence effects in a variety of nonadiabatic processes.  Section \ref{sec:conclusions} contains our conclusions.  The conclusion section outlines the strengths and weaknesses of the bohmion method revealed in the present investigation and outlines several directions for improvement of its capabilities.

\section{The bohmion method in nonadiabatic dynamics\label{sec:Bohmiontheory}}

\subsection{Exact wavefunction factorization\label{sec:EFHyd}}
In this section, we {briefly summarize the specific aspects of} the hydrodynamic formulation of the exact factorization representation \eqref{ExFact} {that we will need in} the subsequent discussions.

Without loss of generality, here we restrict to consider the case of a three-dimensional nuclear coordinate $\boldsymbol{r}$ and a three-dimensional electronic coordinate $\boldsymbol{x}$. The extension to several nuclei and electrons is straightforward. As usual, the Hamiltonian operator $\widehat{H}=\widehat{T}_n+ \widehat{H}_e$ is written as the sum of the nuclear kinetic energy $\widehat{T}_n=-M^{-1}\hbar^2\Delta_{\boldsymbol{r}}/2$ and the electronic Hamiltonian $\widehat{H}_e=\widehat{H}_e(\boldsymbol{r})$ containing the interaction terms. Upon using the Madelung transform  $\chi=\sqrt{D} e^{iS/\hbar}$ and introducing the celebrated quantum potential
$V_Q={-}({M^{-1}\hbar^2}/{2}){\Delta\sqrt{D}}/{\sqrt{D}}$, one writes nuclear dynamics in the hydrodynamic form
\begin{align}
&M(  \partial_t + \bu\cdot \nabla)\bu =- \nabla(V_Q + \epsilon) - \boldsymbol{E}   - \bu \times \boldsymbol{B}\label{EFEPELu1} 
\,,\\
 & \partial_t D + \text{div}(D\bu) = 0\,.\label{Dens-eq}
\end{align}
Here, all differential operators are defined on the nuclear coordinate space  and  the  notation is as follows: $\boldsymbol{A}=\langle\phi|-i\hbar \nabla\phi\rangle$ is the Berry connection with curvature $\boldsymbol{B}=\operatorname{curl}\boldsymbol{A}$, while $\bu=M^{-1}(\nabla S+\boldsymbol{A})$ is the hydrodynamic velocity.  {Also,} $\epsilon$ is the effective electronic potential
\[
   \epsilon(\phi,\nabla\phi):=\braket{\phi|\widehat{H}_e\phi} 
    +\frac{\hbar^2}{2M}\|\nabla\phi\|^2 - \frac{|\boldsymbol{A}|^2}{2M}
    \,,
\]
where we have used the notation $\langle\phi_1|\phi_2\rangle=\int\!\phi^*_1\phi_2\,\de^3 x$ and $\|\phi_1\|^2=\int|\phi_1|^2\de^3 x$. Finally, $\boldsymbol{E}=-\partial_t\boldsymbol{A}-\nabla\langle\phi|i\hbar \partial_t\phi\rangle$. {Then, the {electronic} equation can be written as follows:
\beq\label{sch-e1}
i\hbar\partial_t\phi+i\hbar(\boldsymbol{u}-M^{-1}\boldsymbol{A})\cdot\nabla\phi=\widehat{H}_e\phi-\frac{\hbar^2}{2MD}\operatorname{div}\left(D\nabla\phi\right)+\lambda\phi
\,,
\eeq
where $\lambda(\boldsymbol{r},t)$ is a function  depending on the gauge choice for  $\langle\phi|i\hbar \partial_t\phi\rangle$}. Equations \eqref{EFEPELu1}, \eqref{Dens-eq}, and \eqref{sch-e1} comprise the hydrodynamic formulation of the exact factorization system in \cite{FoHoTr19,Suzuki16}. Combined with the Born-Huang expansion, this system is the basis for the new {\it coupled-trajectory mixed-quantum-classical method} (CT-MQC) in nonadiabatic molecular dynamics  \cite{CoupledvsIndependent}. 

\subsection{The electronic density matrix}
Before introducing the variational structure, we choose to rewrite the system \eqref{EFEPELu1}, \eqref{Dens-eq}, and \eqref{sch-e1} in a slightly different form. First, {after some algebraic manipulations \cite{FoHoTr19},} we notice that
\[
\boldsymbol{E}=-\boldsymbol{u}\times\boldsymbol{B}-\nabla\epsilon+\langle\phi|(\nabla\widehat{H}_e)\phi\rangle+\frac1{MD}\operatorname{div}(D \Bbb{T})
\,.
\]

\noindent
Here, $\Bbb{T}=\operatorname{Re}\Bbb{Q}$ denotes the real part of the {\it quantum geometric tensor} $\Bbb{Q}_{jk}=\langle\partial_j\phi|\partial_k\phi\rangle-\hbar^{-2}A_jA_k$ \cite{Provost80}. By using {the relation above}, equation \eqref{EFEPELu1} becomes
\begin{align}
&M(  \partial_t + \bu\cdot \nabla)\bu =- \nabla V_Q -\langle\phi|(\nabla\widehat{H}_e)\phi\rangle-\frac1{MD}\operatorname{div}(D \Bbb{T})
\,.
\label{EFEPELu2} 
\end{align}
Also, we notice that the electronic equation \eqref{sch-e1} can be rewritten as
\[
i\hbar\partial_t\phi+i\hbar\boldsymbol{u}\cdot\nabla\phi=\widehat{H}_e\phi+\frac{\hbar^2}{4MD}\frac{\delta F}{\delta \phi}+\lambda\phi
\,.
\]
{In the above we have introduced $F=\int D\operatorname{Tr} \Bbb{T}\,\de^3r$, where $\operatorname{Tr}$ denotes the matrix trace. Then, upon writing $T=\operatorname{Tr} \Bbb{T}$, the functional derivative of $F$ is $\delta F/\delta\phi=D\partial T/\partial\phi-\operatorname{div}(D\partial T/\partial\nabla\phi)$}. 

At this point,  we use the density matrix $\rho(\boldsymbol{x},\boldsymbol{x}',t;\br)=\phi(\boldsymbol{x},t;\br)\phi(\boldsymbol{x}',t;\br)^*$ to write $\Bbb{T}_{jk}=\langle\partial_j\rho|\partial_k\rho\rangle$ and $F=\int D\|\nabla\rho\|^2/2\,\de^3r$ where we have used the notation $\langle \rho_1|\rho_2\rangle=\int\rho_1(\boldsymbol{x}',\boldsymbol{x})^*$ $\rho_2(\boldsymbol{x},\boldsymbol{x}')\,\de^3 x\,\de^3 x'$ and $\|\nabla\rho\|^{2}=\langle \partial_k\rho|\partial_k\rho\rangle$. Then, we notice that the chain rule ensures $\delta F/\delta\phi=2(\delta F/\delta\rho)\phi$ so that the {electronic} equation \eqref{sch-e1} can be written as the quantum Liouville equation
\beq\label{el-lio}
i\hbar\left(\frac{\partial}{\partial t}+\boldsymbol{u}\cdot\nabla\right)\rho+\big[\rho,\widehat{H}_e\big]=\frac{\hbar^2}{2MD}\left[\frac{\delta F}{\delta \rho},\rho\right]=\frac{\hbar^2}{2MD}\operatorname{div}\left(D\big[\rho,\nabla\rho\big]\right)
\,,
\eeq
where we have used $\delta F/\delta\rho=-\operatorname{div}(D\nabla\rho)$ and we have applied the Leibniz rule. {Here, we notice the hydrodynamic material derivative ${\partial_t}+\boldsymbol{u}\cdot\nabla$ on the left-hand side, indicating that the electronic evolution is swept by the nuclear flow acting on the nuclear coordinates, which in turn appear parametrically in the unitary propagator of the electronic quantum dynamics; see \cite{FoHoTr19} for further discussions. In addition, we notice} the emergence of the quantity $[\rho,\nabla\rho]$: as recognized in \cite{Mead}, this is a type of    non-Abelian gauge connection. See \cite{Foskett} for recent advances on the appearance of non-Abelian gauge connections in nonadiabatic dynamics. For  later convenience, we introduce the variable $\tilde\rho=D\rho$. In terms of $\tilde\rho$, the equations of motion become 
\begin{align}
&MD(  \partial_t + \bu\cdot \nabla)\bu ={D\nabla V_Q} -\langle\tilde\rho|\nabla\widehat{H}_e\rangle+\frac{\hbar^2}{2M}\partial_j\langle\tilde\rho,\nabla(D^{-1}\partial_j\tilde\rho)\rangle
\label{EFEPELu3} 
\,,\\
 & \partial_t D + \text{div}(D\bu) = 0\label{Dens-eq2}
 \,,\\
 & 
i\hbar \partial_t \tilde\rho + i\hbar\text{div}(\tilde\rho\bu) = \big[\widehat{H}_e,\tilde\rho\big]+
 \frac{\hbar^2}{2M}\operatorname{div}\big(D^{-1}\big[\tilde\rho,\nabla\tilde\rho\big]\big)\,.
 \label{sch-e2}
\end{align}
These nonadiabatic quantum hydrodynamics equations were shown in \cite{FoHoTr19} to possess both a Hamiltonian and variational formulation. The latter is particularly useful in applications of the bohmion method to be discussed later. {We remark that the construction of hydrodyamic models for a nuclear flow interacting with an electronic subsystem also appears in the chemistry literature \cite{Bo11} in the context of mixed quantum-classical dynamics.}

\subsection{Variational structure}
In order to prepare the framework for the formulation of the bohmion method, here we illustrate the variational structure of the hydrodynamic formulation of the exact factorization system. This will be a basic ingredient for introducing the bohmions in the next section.
The Euler-Poincar\'e variational principle $\delta\int_{t_1}^{t_2}\ell\,\de t=0$ for nonadiabatic quantum hydrodynamics involves the Lagrangian
\begin{equation}
  \ell(\bu, D, \xi, \rho) =  \int \bigg[\frac{1}{2}MD |\bu|^2-\frac{\hbar^2}{8M}\frac{|\nabla  D|^2}{D} 
  \\+ \braket{\tilde\rho|i\hbar\xi-\widehat{H}_e} - \frac{\hbar^2D}{4M}\bigg\|\nabla\bigg(\frac{\tilde\rho}{D}\bigg)\bigg\|^2\bigg]\text{d}^3r
 \,. \label{EFHydroL3}
\end{equation}
Here, $\xi(\br,t)$ is the generator of the quantum electronic motion, which will be treated later  in this section. First, we focus on the nuclear hydrodynamic quantities.
The Lagrange-to-Euler map for the nuclear density $D(\boldsymbol{r},t)$ may be written in terms of its initial condition $D_0(\boldsymbol{r}_0)$ by 
\beq
D(\boldsymbol{r},t)=\int D_0(\boldsymbol{r}_0)\delta(\boldsymbol{r}-{\boldsymbol{\eta}}(\boldsymbol{r}_0,t))\,\de^3 r_0\,.
\label{LtEmap}
\eeq
Taking the time derivative of the Lagrange-to-Euler map in \eqref{LtEmap} then recovers the density transport equation in \eqref{Dens-eq2}.
The Lagrangian fluid map $\boldsymbol{\eta}$ in \eqref{LtEmap} plays a crucial role in the hydrodynamic interpretation of equations \eqref{EFEPELu3}-\eqref{sch-e2}. In fact,  
the hydrodynamic velocity $\boldsymbol{u}(\boldsymbol{r},t)$  is defined as the tangent vector to the Bohmian trajectory $\boldsymbol{\eta}(\boldsymbol{r}_0,t)$ given by 
\beq
\dot{\boldsymbol{\eta}}(\boldsymbol{r}_0,t) :=\partial_t {\boldsymbol{\eta}}(\boldsymbol{r}_0,t) 
=\boldsymbol{u}(\boldsymbol{\eta}(\boldsymbol{r}_0,t),t)\,.
\label{veldef}
\eeq
Thus, the Bohmian trajectory  identifies the evolution of Lagrangian fluid parcels labelled by their  initial nuclear position $\boldsymbol{r}_0$ and moving with velocity $\boldsymbol{u}(\boldsymbol{\eta}(\boldsymbol{r}_0,t),t)$. the variations $\delta D$ and $\delta \bu$ arise from the relations \eqref{LtEmap}-\eqref{veldef}. Upon composing \eqref{veldef} by the inverse variable $\boldsymbol{\eta}^{-1}$, the resulting relation $\boldsymbol{u}(\br,t)=\dot{\boldsymbol{\eta}}(\br_0,t)|_{\br_0=\boldsymbol{\eta}^{-1}(\br,t)}$ leads to the variational relations 
\beq
\delta\bu=\partial_t\boldsymbol{w}+(\boldsymbol{u}\cdot\nabla)\boldsymbol{w}-(\boldsymbol{w}\cdot\nabla)\boldsymbol{u}
\,,\qquad\qquad\ 
\delta D=-\operatorname{div}(D\boldsymbol{w})
\,.
\label{vars1}
\eeq
Here, we have introduced $\boldsymbol{w}(\br,t)=\delta{\boldsymbol{\eta}}(\br_0,t)|_{\br_0=\boldsymbol{\eta}^{-1}(\br,t)}$ while the variation $\delta D$ follows from  \eqref{LtEmap}. The  reduction from Lagrangian/Bohmian variables to Eulerian variables in Hamilton's principle for ideal fluid dynamics is called Euler-Poincar\'e reduction \cite{HoMaRa98}. See \cite{FoTr20} for an extension to include the presence of hydrodynamic vortices in QHD.

The presence of the density matrix in the variational principle associated to \eqref{EFHydroL3} is treated here by using the techniques first developed in \cite{BLTr15,Tr19}. In this case, the evolution of the density matrix density $\tilde\rho$ requires some discussion. Usually, the quantum density matrix evolves according to $\rho_0\mapsto U(t)\rho_0U(t)^\dagger$, where $U(t)$ is the unitary propagator. In the present case, we recall that $\rho(\br,t)$ retains parametric dependence on the nuclear coordinates and thus so does the unitary propagator, which we shall denote by $U(\br,t)$. In addition, the electronic density matrix $\rho$ evolves in the frame of the nuclear fluid as indicated by the convective time derivative in the left-hand side of \eqref{el-lio}. Then, the density matrix density $\tilde\rho=D\rho$ evolves according to \cite{FoHoTr19}
\beq
\tilde\rho=
\int\! \hat\rho(\br_0,t)\,\delta(r-{\eta}(\br_0,t))\,\de^3 r_0
\,,
\qquad\text{with} \qquad
\hat\rho(\br_0,t)=U(\br_0,t)\tilde\rho_0(\br_0)U^\dagger(\br_0,t)
\,.
\label{dmd-evol}
\eeq
In terms of these variables, the quantum generator of motion $\xi(\br,t)$ is defined as
$\xi(\br,t)=\dot{U}(\br_0,t){U}(\br_0,t)^\dagger|_{\br_0=\boldsymbol{\eta}^{-1}(\br,t)}$,
so that equation \eqref{sch-e2} has the general structure $ \partial_t \tilde\rho + \text{div}(\tilde\rho\bu) = \big[\xi,\tilde\rho\big]$. Also, upon denoting $\nu(\br,t)=\delta{U}(\br_0,t){U}(\br_0,t)^\dagger|_{\br_0=\boldsymbol{\eta}^{-1}(\br,t)}$, one obtains the variational relations
\beq
\delta\tilde\rho=[\nu,\tilde\rho]-\operatorname{div}(\boldsymbol{w}\tilde\rho)
\,,\qquad\qquad\ 
\delta \xi = \partial_t \nu -\boldsymbol{w}\cdot\nabla\xi + \bu\cdot\nabla\nu - [\xi,\nu]
\,,
\label{vars2}
\eeq
while $\boldsymbol{w}$ is given as in \eqref{vars1}.

\subsection{Regularization and the bohmion method\label{BohmionMeth}}


Of course, setting $\hbar\to 0$ would eliminate the gradient terms in \eqref{EFHydroL3}, thereby allowing for Young measure ($\delta$-function) solutions in the variables $D$ and $\tilde\rho$. This procedure is based on a singular weak limit leading to the mean-field model and eliminating electronic decoherence. Conversely, the singular  solutions cannot  exist for $\hbar\neq0$ due to the structure of the order $O(\hbar^2)$ terms. 

The key idea in the derivation of the bohmion model is to \emph{regularize} the hydrodynamic description of the nuclear variables by performing a spatial smoothing of the order $O(\hbar^2)$ density-gradient terms in \eqref{EFHydroL3}, rather than neglecting these terms. Then, one obtains a dispersive regularization of  nonadiabatic quantum hydrodynamics (RQHD) and restores the $\delta$-function solutions  \emph{without} enforcing $\hbar\to 0$. 

These measure-valued solutions -- called \emph{bohmions} -- may be interpreted as describing statistical ensembles of classical nuclear trajectories. 
The corresponding bohmion equations of motion enhance the underlying mean-field model and extend its range of applicability by enabling it to capture decoherence effects. The capability to capture decoherence thus achieved also extends to the celebrated surface hopping method, which exploits the Born-Huang  surfaces.

More specifically, in the bohmion model the gradient terms in \eqref{EFHydroL3} are \emph{mollified} by a convolution $K(\br-\br')$ which introduces the following regularized quantities  
\begin{equation}
\bar{D}(\br,t)=\int K(\br-\br') D(\br',t)\,\de^3r
\,,\qquad\qquad\ 
\bar{\rho}(\br,t)=\int K(\br-\br') \tilde\rho(\br',t)\,\de^3r
 \,. \label{def: mollifier}
\end{equation}
A similar  approach was recently applied to regularize conical intersections in adiabatic dynamics with geometric phase effects \cite{RaTr20}. The mollifier is typically rotation-invariant and depends on a  lengthscale parameter ${\alpha}$ so that the limit ${\alpha}\to 0$ recovers the original hydrodynamic variable $D$. For example, ${\alpha}$ could be the width of a Gaussian convolution kernel. Then, we consider the following regularized version of the Lagrangian \eqref{EFHydroL3}:
\begin{equation}
  \ell =  \int \bigg[\frac{1}{2}MD |\bu|^2-\frac{\hbar^2}{8M}\frac{(\nabla 
  \bar{D})^2}{\bar{D}} 
  \\+ \braket{\tilde\rho|i\hbar\xi-\widehat{H}_e} - \frac{\hbar^2}{4M}\Big\|\nabla\Big(\frac{\bar\rho}{\bar{D}}\Big)\Big\|^2\bigg]\text{d}^3r
 \,, \label{EFHydroL4}
\end{equation}
so that the associated RQHD equations arise from Hamilton's variational principle upon using \eqref{vars1} and \eqref{vars2}; see \cite{FoHoTr19} for their explicit form.

A remarkable feature of these RQHD equations (which is not shared by the QHD equations) is that for $\hbar\neq0$ they admit singular solutions  in which the nuclear density is given by a finite
train of $\delta$-functions. These $\delta$-functions are called \emph{bohmions} and follow \emph{bohmion
trajectories} in configuration space. In particular,  replacing  the initial condition $
\tilde\rho_0(\br_0)=\sum_{a}w_a\varrho_a^{(0)}\delta(\br_0-\bq_a^{(0)})
$ in \eqref{dmd-evol} leads to \cite{FoHoTr19}
\beq
\tilde\rho(\br,t)=\sum_{a=1}^Nw_a\varrho_a(t)\delta(\br-\bq_a(t))
\,,\qquad
\text{with}
\qquad
\varrho_a(t):={\cal U}_a(t)\varrho_a^{(0)}{\cal U}_a(t)^\dagger
\,.
\label{DMDansatz}
\eeq
Here, we have denoted ${\cal U}_a(t):=U(\bq_a^{(0)},t)$ and we set $\varrho_a^{(0)}(\boldsymbol{x},\boldsymbol{x}')=\varphi_a^{\scriptscriptstyle (0)}(\boldsymbol{x}){\varphi_a^{\scriptscriptstyle (0)}}(\boldsymbol{x}')^*$, so that $\varrho_a(\boldsymbol{x},\boldsymbol{x}',t)=\varphi_a(\boldsymbol{x},t)\varphi_a(\boldsymbol{x}',t)^*$  at all times. {We notice that the ansatz \eqref{DMDansatz} comprises part of a singular momentum map structure which is well known in geometric mechanics \cite{HoMa04,HoTr2009}.}
Then, by using the ansatz 
\beq
D(\br,t)=\sum_{a=1}^Nw_a\delta(\br-\bq_a(t))
\label{BohmAns1}
\eeq and denoting $\xi_a=\dot{\cal U}_a{\cal U}_a^\dagger$, one obtains the nonadiabatic bohmion Lagrangian
\begin{multline}
L(\{\bq\},\{\dot{\bq}\},\{\varrho\})  =\sum_{a} w_a\Bigg(\frac{M}2\dot{\bq}_a^2+\braket{\varrho_a,i\hbar\xi_a-\widehat{H}_e(\bq_a)}
  \\
  + \frac{\hbar^2}{8M} \sum_{b} w_b(1-2\langle\varrho_a |\varrho_b\rangle)  {\int 
  \frac{\nabla K(\br-\bq_a)\cdot\nabla K(\br-\bq_b)}
  {\sum_{c} w_c K(\br-\bq_c)} \,\text{d}^3r }
  \Bigg)
 \,.
 \label{BohmionLagr}
\end{multline}
Each of the bohmions supports an electronic state which
has its own unitary dynamics along the corresponding trajectory. The interactions of a finite number of 
bohmions and their associated electronic states produce a  finite-dimensional trajectory-based closure model that arises from Hamilton's principle $\delta\int_{t_1}^{t_2}L\,\de t=0$. As discussed in \cite{FoHoTr19}, the latter requires the variations
\[
\delta \xi_a = \partial_t \nu_a - [\xi_a,\nu_a]
\,,\qquad\qquad\ 
\delta \varrho_a=[\nu_a,\varrho_a]
\,,
\]
where $\nu_a=(\delta{\cal U}_a){\cal U}_a^\dagger$. These relations are easily verified from the definitions of $\xi_a$ and $\varrho_a$. Eventually, the bohmion motion is governed by the Euler-Lagrange equations for $\bq_a$, which are accompanied by a sequence of quantum Liouville equations for $\varrho_a$. The latter read 
\beq
i\hbar\dot{\varrho}_a = 
\big[\widehat{H}_e(\bq_a),{\varrho}_a\big]
+
\frac{\hbar^2}{2M} \sum_{b}w_b \left[{\varrho_b},{\varrho}_a\right]{ 
\int \frac{\nabla K(\br-\bq_a)\cdot\nabla K(\br-\bq_b)}{\sum_{c} w_c K(\br-\bq_c)} \,\de^3r
}\,.
\label{BohmionDMDeq}
\eeq
Upon writing ${\varrho}_a(\boldsymbol{x},\boldsymbol{x}',t)=\varphi_a(\boldsymbol{x},t)\varphi_a(\boldsymbol{x}',t)^*$, we can also  write the corresponding Schr\"odinger equation as follows:
\beq
i\hbar\dot{\varphi}_a = 
\widehat{H}_e(\bq_a){\varphi}_a
+
\frac{\hbar^2}{M} \sum_{b}w_b \langle\varphi_b|\varphi_a\rangle\varphi_b{ 
\int \frac{\nabla K(\br-\bq_a)\cdot\nabla K(\br-\bq_b)}{\sum_{c} w_c K(\br-\bq_c)} \,\de^3r
}\,.
\label{BohmionDMDeq2}
\eeq

At this point, the problem has been made finite-dimensional and the bohmion motion is governed by the Euler-Lagrange equations for $\bq_a$. We remark that the present treatment is inherently nonadiabatic, {although} there seems to be no clear sense in which certain terms in equation \eqref{BohmionDMDeq2} are particularly responsible for the nonadiabatic coupling terms appearing in Born-Huang expansions. 

 In the $\hbar\to0$ limit the bohmion trajectories  reduce to uncoupled classical trajectories describing a statistical ensemble of nuclei evolving under the mean-field influence of electronic degrees of freedom.   In this sense, the bohmion picture  places classical and quantum trajectories on the same footing, with $\hbar$ playing a transparent role as a coupling constant \cite{FoHoTr19}. Note that the limit $\hbar\to0$ in bohmion dynamics is equivalently achieved by taking the smoothing lengthscale $\alpha\to\infty$, which has the effect of washing out the contributions from the order  $O(\hbar^2)$ terms. In the opposite limit $\alpha\rightarrow0$
we have that $\overline{D}\rightarrow D$ and $\overline{\rho}\rightarrow\rho$
with $\dot{q}_{a}=u\left(q_{a}\right)$ and so formally the trajectories
$\dot{q}_{a}$ approach the exact nuclear Bohmian trajectories.
In the intermediate regime, we see that  the last term in \eqref{BohmionLagr} is essential in that it retains the nonlocal  features occurring in bohmion dynamics, so that the motion of each bohmion affects all the other bohmions.  Moreover, as bohmion dynamics is Hamiltonian, we remark that it naturally inherits conservation of energy and momentum.

To gain insight into the solution properties of the model, in the next section we will explore these properties in more detail by considering a series of numerical benchmark problems.

\section{Results for model systems\label{sec:results}}

In this section we {present the results obtained by testing the bohmion method on four model systems, including the three so-called
\emph{Tully models}, Tully I, II, III. Comparisons are also made with well-established schemes  including mean-field (Ehrenfest), trajectory surface hopping (TSH), and the coupled-trajectory mixed-quantum-classical method (CT-MQC) \cite{CoupledvsIndependent}; see Appendix \ref{app:comp}.

All models considered here are two-state models with a one-dimensional nuclear
coordinate $r$ and molecular Hamiltonian given by
\begin{equation}
\widehat{H}=-\frac{1}{2M}\partial_{r}^{2}+\widehat{H}_{e}\left(r\right)
\end{equation}
featuring the electronic Hamiltonian (in a diabatic basis)
\begin{equation}
\hat{H}_{e}\left(r\right)=\begin{pmatrix}H_{11}\left(r\right) & H_{12}\left(r\right)\\
H_{21}\left(r\right) & H_{22}\left(r\right)
\end{pmatrix}.
\end{equation}
Depending on the explicit form of $\hat{H}_{e}$, t}he Tully models were first introduced in the 90s \cite{Tully} and since then have become a standard testing ground for any new approach to nonadiabatic molecular dynamics. 
These simple two-state models with a one-dimensional nuclear degree of freedom enable exact quantum mechanical simulations to be performed against which approximate schemes may be compared. At the same time, the Tully models can mimic realistic higher-dimensional nonadiabatic molecular processes. For example, parallels can be drawn between Tully I and the photoisomerization of ethylene (as well as many other photodynamical processes), and similar comparisons can be made for the other Tully models \cite{Curchod}.

In each case, we prepare a nuclear wavepacket at spatial
infinity on the lowest BO electronic potential energy surface, then
study what happens as it encounters a region of nonadiabatic coupling. Specifically, we are interested
in whether bohmion dynamics accurately capture BO population transfer
and electronic decoherence.

{Here, we work in atomic units and use the same initial conditions as those considered in \cite{CoupledvsIndependent}. In particular,  we consider the
initial molecular wavefunction
\begin{equation}\label{InMolWF}
\Psi_{0}=\frac{1}{\left(\pi\Delta_{0}^{2}\right)}\exp\left(-\frac{1}{2}\left(\frac{r-r_{0}}{\Delta_{0}}\right)^{2}+ik_{0}r\right)|1\rangle
\end{equation}
where $\Delta_{0}=20/k_{0}$ and where the diabatic electronic basis
have been labelled $\left\{ |1\rangle,|2\rangle\right\} $. Evidently, $r_0$ is the centre of the initial wavepacket while $k_0$ is its momentum; {we will consider different values depending on the case under consideration}. We take
the initial exact nuclear wavefunction to be $\chi_{0}=\exp\left(-\frac{1}{2}\left(r-r_{0}\right)^{2}/\Delta_{0}^{2}+ik_{0}r\right)/\left(\pi\Delta_{0}^{2}\right)$,
leading to an initial hydrodynamic density $D_{0}=\exp\left(-\left(r-r_{0}\right)^{2}/\Delta_{0}^{2}\right)/\left(\Delta_{0}\sqrt{\pi}\right)$
and hydrodynamic velocity $u_{0}=k_{0}/M$ as well as the electronic
density matrix density $\tilde{\rho}_{0}=D_{0}|1\rangle\langle1|$.

To model the initial density, we write 
\begin{equation}
D_{0}=\frac{1}{\Delta_{0}\sqrt{\pi}}\exp\left(-\left(\frac{r-r_{0}}{\Delta_{0}}\right)^{2}\right)\sim\frac{1}{N}\sum_{a=1}^{N}\delta\left(r-q_{a}\left(0\right)\right)
\end{equation}
as a finite train of $\delta$-functions where the initial bohmion
positions $q_{a}\left(0\right)$ are randomly sampled from a normal
distribution with mean $\mu=r_{0}$ and variance $\sigma=\Delta_{0}^{2}/2$.
{Sampling was performed
with a \emph{pseudo}random number generator and also with a \emph{quasi}random
number generator based on an inverse CDF transform of the one-dimensional
Sobol sequence, with both methods giving accurate results. The results
presented here use the quasirandom sampling method, for which we found
faster convergence as the number of trajectories was increased. This
is not surprising: the convergence properties of Monte Carlo and quasi-Monte
Carlo methods are well-studied and the scaling of quasi-Monte Carlo
methods (with numbers of samples, but also with dimensionality \cite{Shalashilin sobol})
is known to be superior, at least asymptotically.}

The initial bohmion velocities are $\dot{q}_{a}\left(0\right)=k_{0}/M$
and the initial electronic density matrices are $\varrho_{a}\left(0\right)=|1\rangle\langle1|$.
We numerically integrate the bohmion equations with these initial
conditions, using an RK4 scheme with a step size of $0.5$ and take
$M=2000$ (which is comparable to the proton mass in atomic units).
At each time step, integrals which appear on the RHS of the bohmion
equations must be evaluated over nuclear coordinate space. We find
a simple trapezoidal rule using a sample spacing of $\alpha/3$ is
adequate.

The quantities of our particular interest are the BO populations
{
\begin{equation}
P_{j}\left(t\right)=\frac{1}{N}\sum_{a=1}^{N}P_{ja}\left(t\right)=\frac{1}{N}\sum_{a}\mathrm{Tr}\left[\pi_{j}\left(q_{a}\left(t\right)\right)\varrho_{a}\left(t\right)\right] ,
\end{equation}
where $\pi_{j}=|j\rangle\langle{j}|$  is the projection  onto the lower ($j=1$) or 
upper ($j=2$) BO state}. We are also interested in the coherence measure
\begin{equation}
C\left(t\right)=\frac{1}{N}\sum_{a}P_{1a}\left(t\right)P_{2a}\left(t\right).
\end{equation}
In this last quantity, the contribution $P_{1a}\left(t\right)P_{2a}\left(t\right)$
from the $a$th bohmion goes to zero when the electronic density matrix
$\varrho_{a}$ associated with the trajectory tends to either the
lower BO state (in which case $P_{2a}=0$) or upper BO state (in which
case $P_{1a}=0$). The decay of this quantity away from regions of
nonadiabaticity is therefore an indicator of electronic decoherence.
}

\subsection{Tully I (single avoided crossing)}
\begin{figure}[!h]
\begin{centering}
\includegraphics[width=0.5\paperwidth]{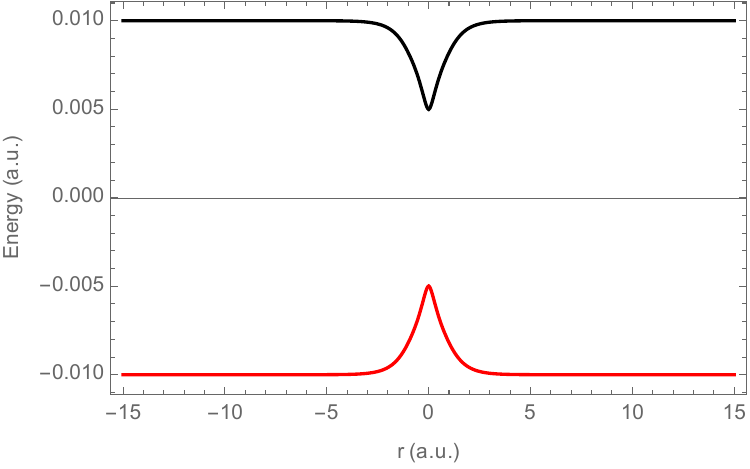}
\par\end{centering}
\caption{Tully I BO energy surfaces.\label{fig:TullyIBO}}
\end{figure}

\begin{figure}[!h]
\begin{centering}
\includegraphics[width=0.8\paperwidth]{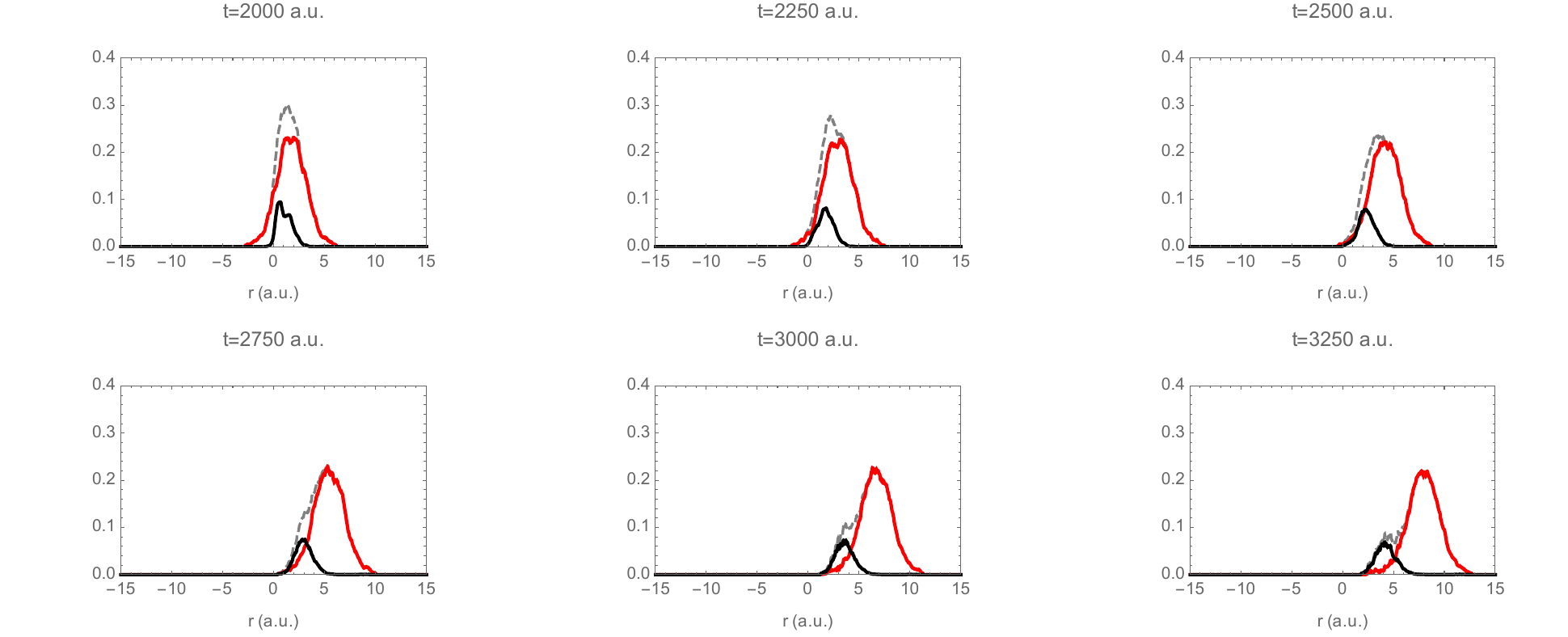}
\par\end{centering}
\caption{Wavepacket splitting in Tully I. $k=10$, ${N}=4000$, ${\alpha}=1/20$. In each snapshot, the dotted line represents the regularized probability density  $\bar{D}(r,t)=\int K(r-r') D(r',t)\,\de r$. The red and black lines represent the populations of the lower and upper BO surfaces respectively, which are computed from the regularized density matrix density $\bar{\rho}(r,t)=\int K(r-r') \tilde\rho(r',t)\,\de r$.\label{fig:TullyIdens}}
\end{figure}

Tully I is defined by the electronic matrix elements 
\begin{align}
H_{11}\left(r\right)&=a\left[1-\exp\left({-br}\right)\right],\quad r>0,
\\
H_{11}\left(r\right)&=-a\left[1-\exp\left({br}\right)\right],\quad r<0,
\\
H_{22}\left(r\right)&=-H_{11}\left(r\right),
\\
H_{12}\left(r\right)&=H_{21}\left(r\right)=c\exp\left({-dr^{2}}\right)
\end{align}
with $a=0.01,b=1.6,c=0.005,d=1.0$. The BO energy surfaces are illustrated
in Figure \ref{fig:TullyIBO}. 
Note that there is a single avoided crossing, centred
at $r=0$. When the nuclear wavepacket (coming in from spatial infinity on the lowest BO electronic state) encounters this avoided crossing, some nonadiabatic transitions into the upper BO state occur. The wavepacket then branches (see Figure \ref{fig:TullyIdens}), with the lower BO wavepacket moving faster than the upper BO wavepacket.

Recall that the bohmion method involves the introduction of a mollifier, as in \eqref{def: mollifier}. Here, we {take the mollifier} to be a Gaussian filter (with some width ${\alpha}$) in all of our simulations. In general, one expects the accuracy of the method to improve as ${\alpha}\to 0$, though in practice this {would require additional} bohmions (larger ${N}$) to achieve reasonable convergence of the results. It should be noted that another difficulty in taking ${\alpha}$ to be very small can arise, as follows. To understand this difficulty, recall that the regularized quantum potential represents a non-local interaction potential for the bohmions which has characteristic energy scale ${\cal E}={\hbar^2}/{M {\alpha}^2}$, as can be seen from the second line of \eqref{BohmionLagr} when $K$ is taken to be a Gaussian of width ${\alpha}$. Consequently, for small ${\alpha}$, we expect the electronic density matrix elements to oscillate with frequency $\omega\sim{\hbar}/{M {\alpha}^2}$ whose growth as ${\alpha}^{-2}$, can impose very small timestep requirements in our numerical algorithm. In each plot we indicate our final choice for ${\alpha}$ and ${N}$. See Figure \ref{dependence} in Section \ref{sec:TullyII} and Figure \ref{archdependence} in Section \ref{sec:DAmodel} for the dependence of the results on ${\alpha}$.
\begin{figure}[!h]
\smallskip
\begin{centering}
\includegraphics[width=0.35\paperwidth]{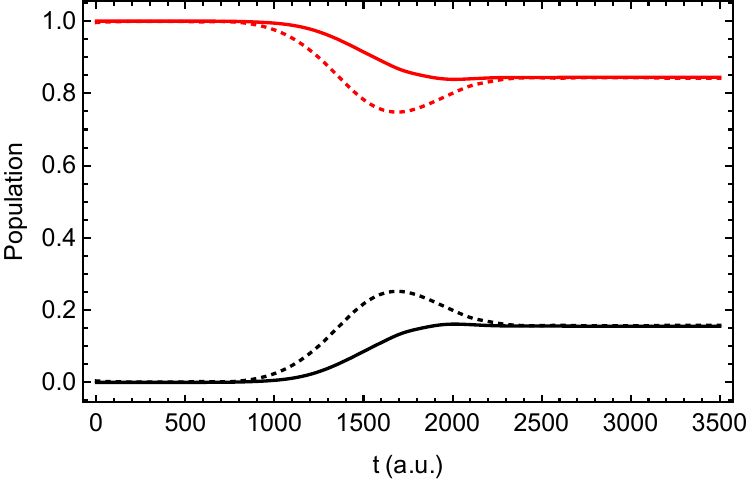}\includegraphics[width=0.35\paperwidth]{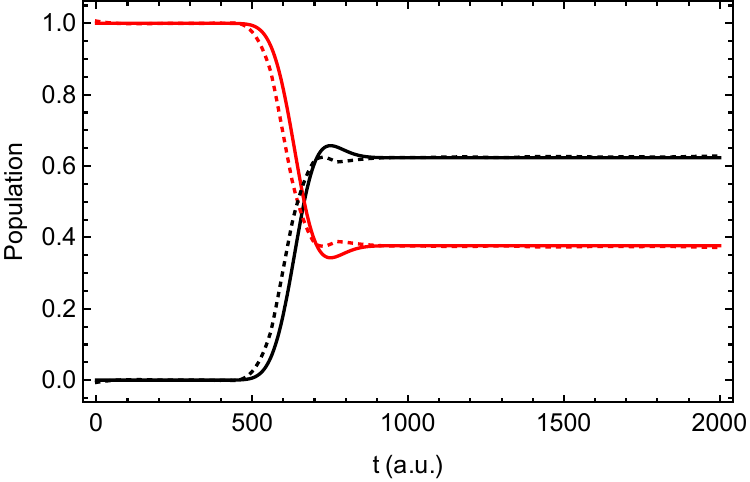}
\includegraphics[width=0.35\paperwidth]{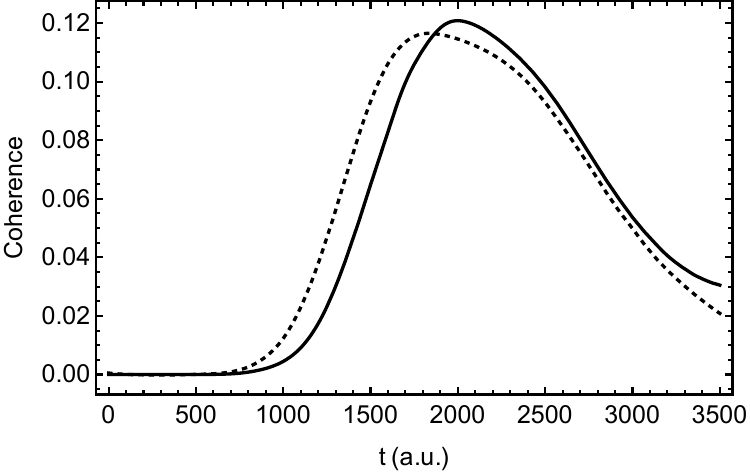}\includegraphics[width=0.35\paperwidth]{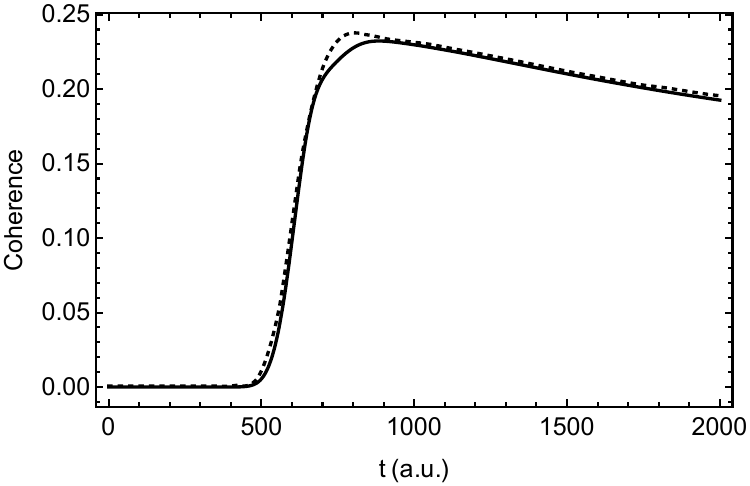}
\par\end{centering}
\caption{BO populations and coherence for Tully I. Left: $k=10$, ${N}=4000$, ${\alpha}=1/20$. Right: $k=25$, ${N}=1000$, ${\alpha}=1/20$. The dotted lines are  the exact results for comparison. \label{fig:TullyIResults}}
\end{figure} 
\paragraph{Decoherence.}
We show two simulations for Tully I, with the nuclear wavepacket given initial
momenta $k=10$ and $k=25$ respectively. {The centre of the initial wavepacket in \eqref{InMolWF} is set to $r_{0}=-8$ a.u. in both cases.} In Figure \ref{fig:TullyIResults} we plot the
BO populations and coherence measure in both cases. 
For the choice ${\alpha}=1/20$, the plots are already in good agreement with exact results; see Figure \ref{fig:Grossplots} in Appendix \ref{app:comp}. 
Particularly noteworthy is the accurate description of decoherence, a phenomenon
which many traditional methods such as TSH and Ehrenfest miss for this model. {However, the CT-MQC method does capture decoherence to some extent \cite{CoupledvsIndependent}}.

\paragraph{Nuclear wavepacket splitting.}
Another important effect accurately captured by the bohm-ion method is nuclear wavepacket splitting. In Figure \ref{fig:TullyIdens} we show snapshots
of the nuclear density and BO projections during the course of the $k=10$ simulation. 
We see that the nuclear wavepacket ultimately splits into two
wavepackets, one located on the lower BO surface and one on the higher
BO surface. The latter wavepacket moves slower than the former, and
so the two move apart as one would expect. This behaviour {would be} missed by schemes based on independent nuclear trajectories. For example, in Ehrenfest
dynamics the ultimate fate of a trajectory is that it follows a potential
energy surface which is some weighted average of the two BO surfaces
(rather than one or the other), and so nuclear wavepacket splitting of this sort is impossible \cite{Tully}. Thus, the coupling between the bohmion trajectories, through the (regularized) quantum potential, is therefore {instrumental} in capturing this effect. 

\subsection{Tully II (dual avoided crossing)\label{sec:TullyII}}
\begin{figure}[h!]
\smallskip
\begin{centering}
\includegraphics[width=0.5\paperwidth]{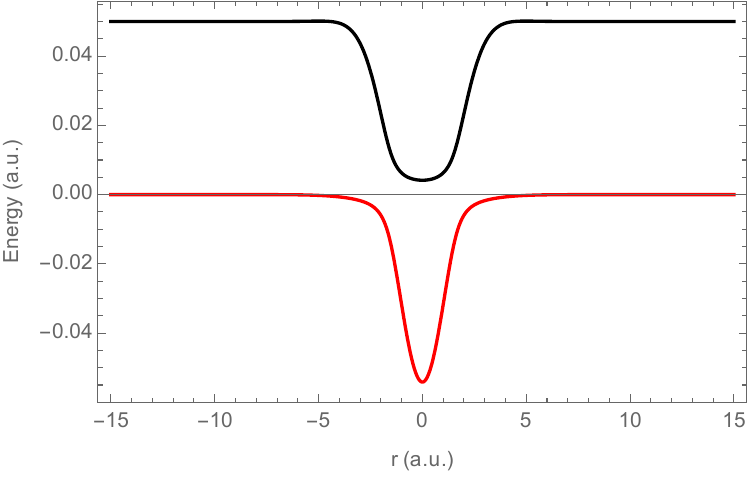}
\par\end{centering}
\caption{Tully II BO energy surfaces.\label{fig:TullyIIBO}}
\end{figure}

Tully II is defined by the electronic matrix elements 
\begin{align}
H_{11}\left(r\right)&=0,
\\
H_{22}\left(r\right)&=-a\exp\left({-br^2}\right)+e_{0},
\\
H_{12}\left(r\right)&=H_{21}\left(r\right)=c\exp\left(-dr^{2}\right)
\end{align}
with $a=0.1,b=0.28,c=0.015,d=0.06,e_{0}=0.05$. The BO energy surfaces
are illustrated in Figure \ref{fig:TullyIIBO}. 
Note that there are two avoided crossings,
located either side of $r=0$. When the nuclear wavepacket (coming in from spatial infinity on the lowest BO electronic state) encounters the first avoided crossing, some nonadiabatic transitions into the upper BO state occur and the wavepacket branches as in Tully I. The wavepackets then encounter the second avoided crossing at which further transitions occur. The wavepackets can recombine at this point leading to interference effects whose strength  depends on the momentum of the 
\begin{figure}[h!]
\smallskip
\begin{centering}
\includegraphics[width=0.35\paperwidth]{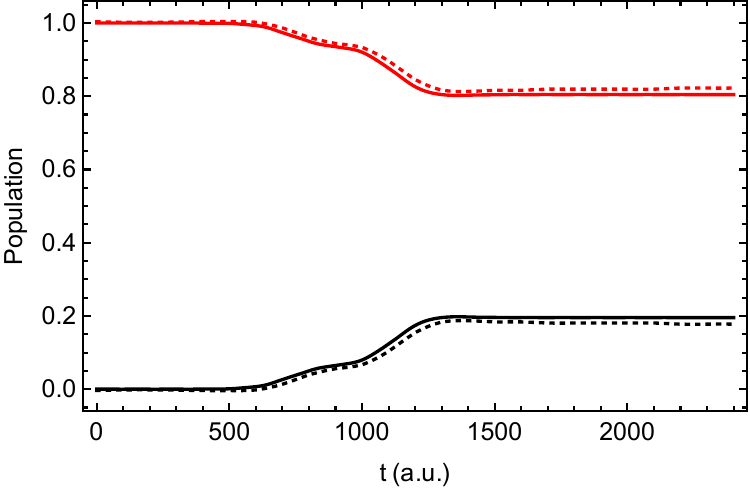}\includegraphics[width=0.35\paperwidth]{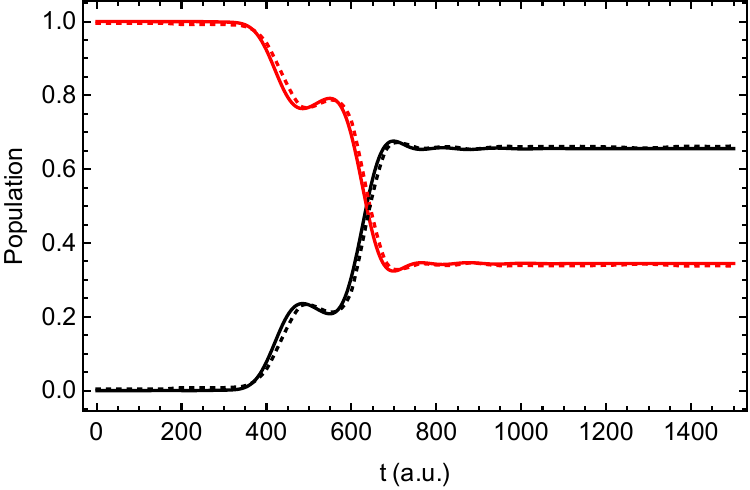}
\includegraphics[width=0.35\paperwidth]{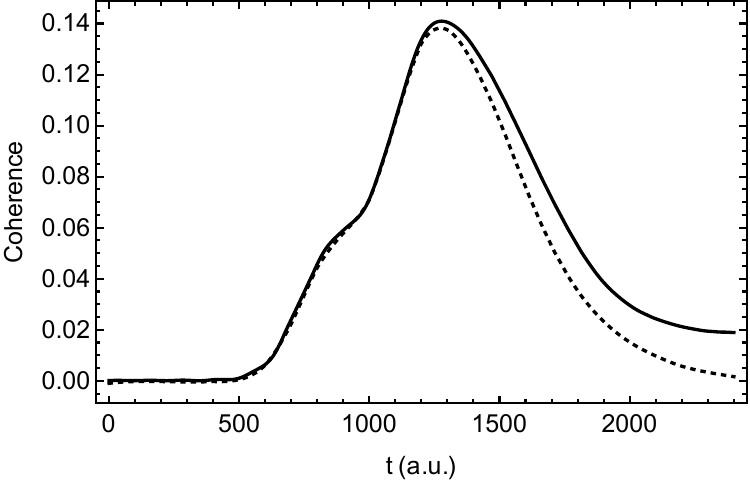}\includegraphics[width=0.35\paperwidth]{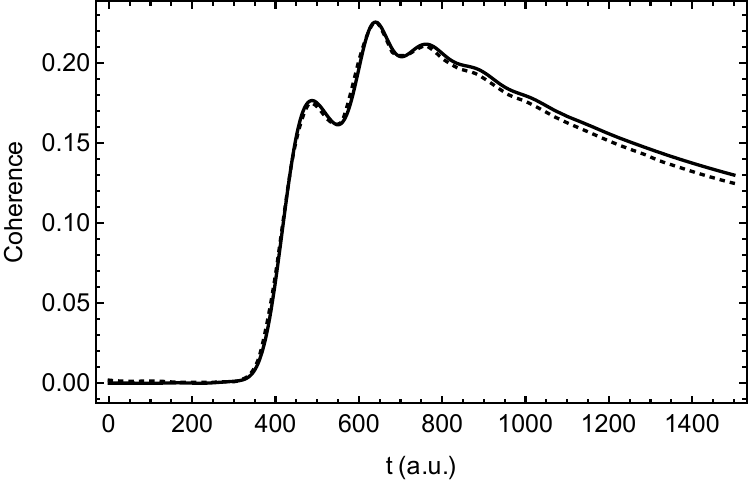}
\par\end{centering}
\caption{BO populations and coherence for Tully II. Left: $k=16$, ${N}=4000$,
${\alpha}=1/20$. Right: $k=30$, ${N}=1000$, ${\alpha}=1/20$.  The dotted lines are  the exact results for comparison. \label{fig:TullyIIResults}}
\end{figure}
initial wavepacket \cite{Tully}. It was recently pointed out that a molecular analogue of Tully II, whose dynamics are characterized by multiple crossings between electronic states, can be found in the photodynamics of the molecule DMABN \cite{Curchod}.
\paragraph{Decoherence.}
We show two simulations, with the nuclear wavepacket given initial
momenta $k=16$ and $k=30$ respectively, {while we set again 
 $r_{0}=-8$ a.u. in \eqref{InMolWF}}. In Figure \ref{fig:TullyIIResults} we
plot the BO populations and coherence measure in both cases. 
Once again, upon comparing with Figure \ref{fig:Grossplots} in Appendix \ref{app:comp}, the decoherence is captured particularly well by the bohmions.
Another impressive feature of the bohmion dynamics is the population
transfer for the lower momentum ($k=16$) scattering. The bohmion
model captures the final BO populations with high accuracy.

\paragraph{Trajectories of bohmions.}
{In order to present a more detailed discussion of the results, here we present some of the specific dynamical features that were obtained for the Tully II model with $k_0=30$. 
 
An attractive feature of quantum trajectory approaches is that 
{one can easily visualise} the dynamics. In Figure \ref{DAC} (left-hand plot) we display
$20$ representative bohmion trajectories from a simulation with $\alpha=1/10$,
$N=100$. The trajectories in this plot are coloured in
a way which indicates the BO population calculated from the associated
electronic density matrix. We see a clear splitting of the trajectory
ensemble after passage through the avoided crossings, into a collection
of blue trajectories (upper BO state) and yellow trajectories (lower
BO state). These two branches of trajectories continue to separate
further from one another, with the trajectories on the upper BO surface
moving slower than those on the lower BO surface as we would expect. 
\begin{figure}[h!]
\begin{centering}
\includegraphics[scale=0.47]{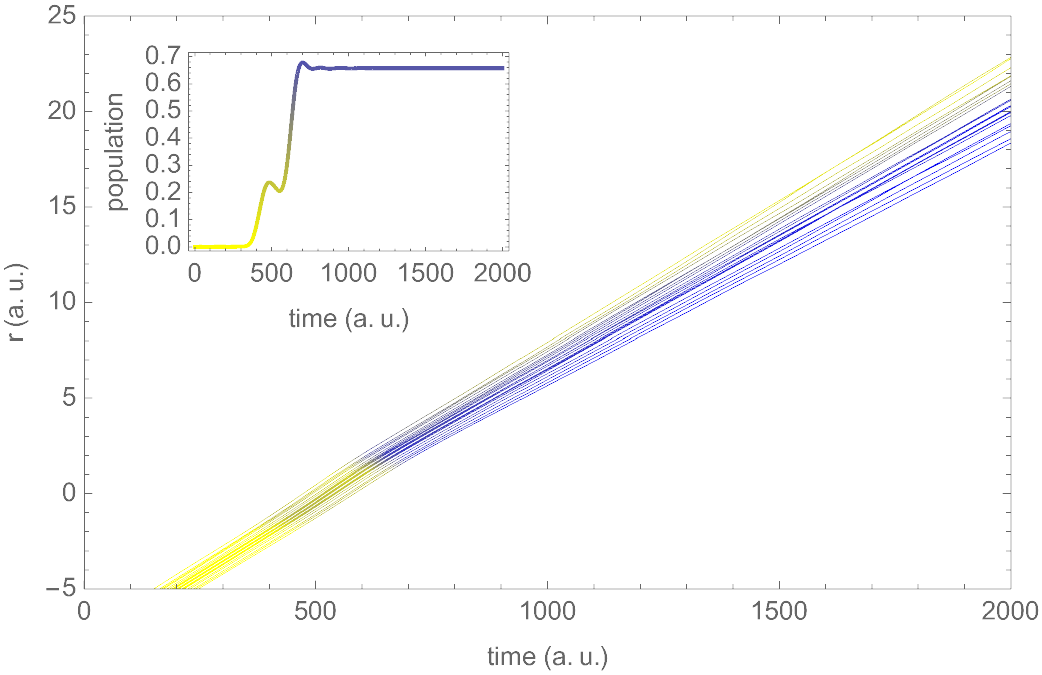}\includegraphics[scale=0.47]{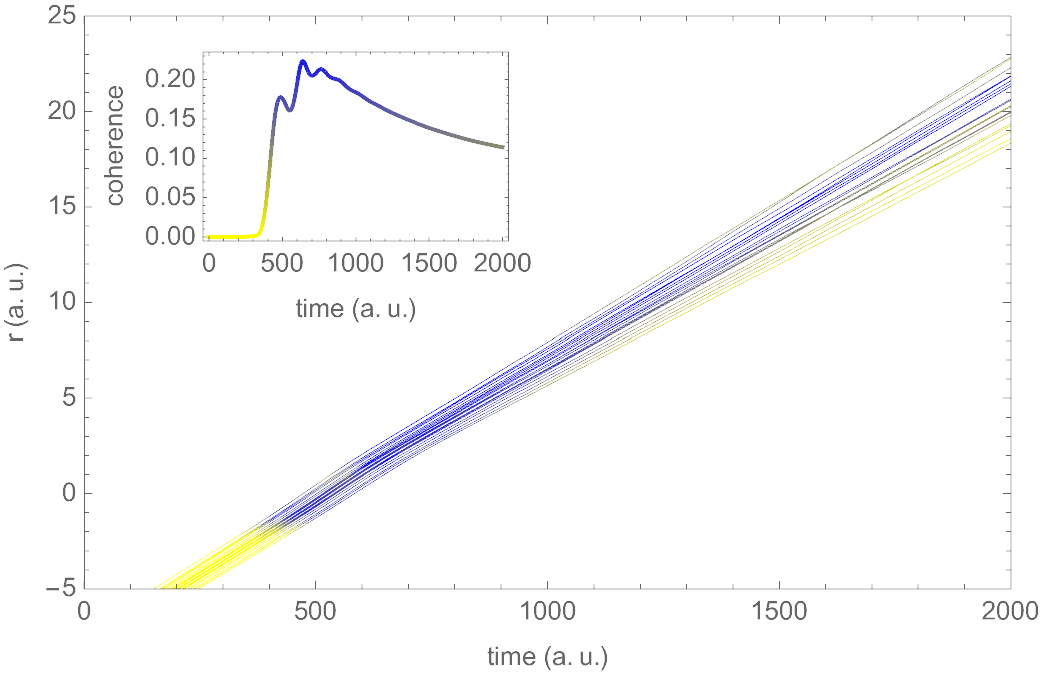}
\par\end{centering}
\caption{Left: population dynamics for $\alpha=1/10$. Note the splitting of the
wavepacket into two, corresponding to the lower BO state (yelow) and
the upper BO state (blue). Right: coherence dynamics for $\alpha=1/10$. Following the nonadiabatic
events at the avoided crossings ($t>1000$), the gradual appearance
of yellow indicates that the corresponding trajectories have decohered.
These are to be contrasted with the trajectories in the Ehrenfest limit (Figure \ref{ehrlimit}).}
\label{DAC}
\end{figure}

An alternative way of colouring these same trajectories is given in the right-hand plot in
Figure \ref{DAC}. Here, the colouring indicates the coherence measure contribution
$P_{1a}\left(t\right)P_{2a}\left(t\right)$ of the corresponding trajectory
$q_{a}\left(t\right)$. After the passage through the avoided crossing
($t>1000$) we see the gradual appearance of yellow which indicates
that the corresponding trajectory has decohered, i.e. the electronic
state has settled into either the lower or upper BO state. Blue, on
the other hand, means that the state is a superposition of the two. 
\begin{figure}[h]
\begin{centering}
\includegraphics[scale=0.47]{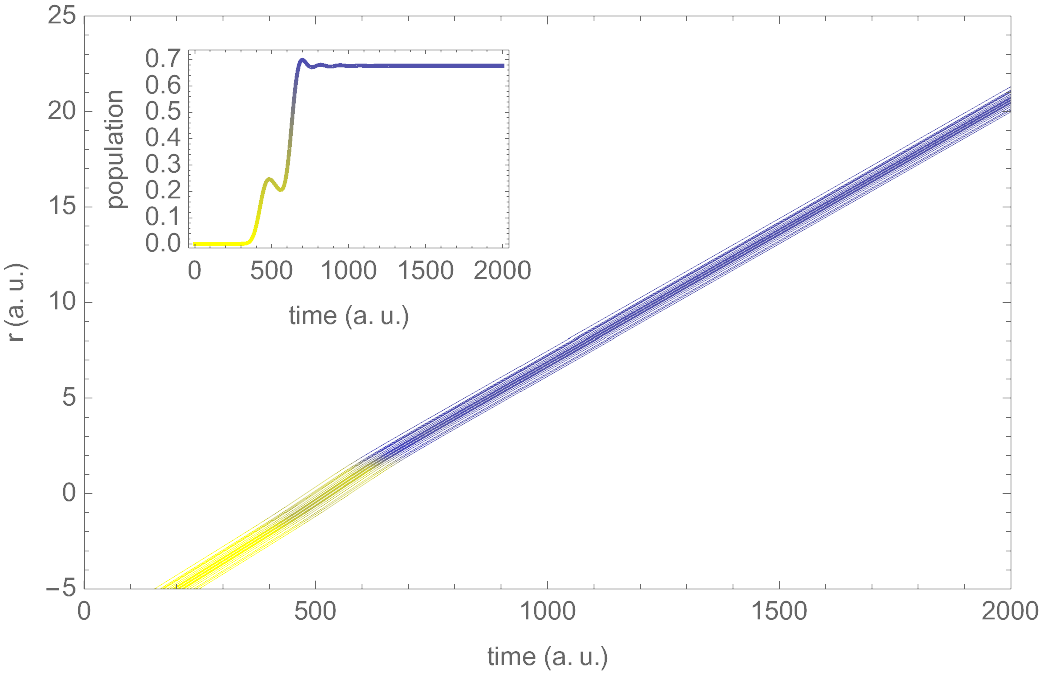}\includegraphics[scale=0.47]{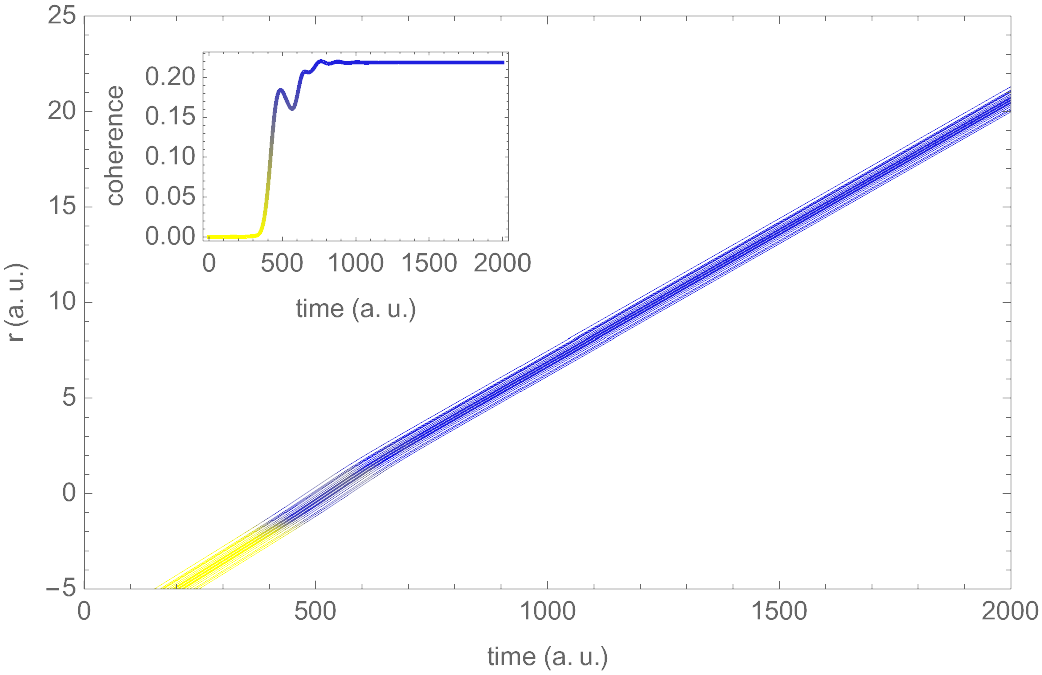}
\par\end{centering}
\caption{Left:
population dynamics for $\alpha=10$ (Ehrenfest limit). Right: coherence dynamics for $\alpha=10$ (Ehrenfest limit).}
\label{ehrlimit}
\end{figure}

The behaviour of these bohmion trajectories is to be contrasted with that of trajectories in the $\hbar\to0$ limit, which is equivalent to taking $\alpha$ large (see Section \ref{sec:Bohmiontheory}). Neither of the effects just described are
captured in this limit: there is no splitting of the
wavepacket and no electronic decoherence occurs once the trajectories
have passed through the avoided crossings, as can be seen in Figure
\ref{ehrlimit} (where we take $\alpha=10$). These effects are due to the $\hbar$-induced couplings of particle histories in the bohmion equations
, all of which have been washed out
by the spatial smoothing.

The choice of the number of bohmions $N$ and smoothing
lengthscale $\alpha$ deserves further discussion. Figure \ref{dependence} shows
the coherence dynamics for a fixed smoothing lengthscale $\alpha=1/10$,
and $N=25,50,100$ bohmions, showing clear convergence with
increasing $N$. We see that, for this model and choice
of initial conditions, good convergence is obtained with as few as
$N=50$ bohmions. In Figure \ref{dependence} we also show converged results
for varying choice of lengthscale $\alpha=10,1/5,1/10,1/20$. 
\begin{figure}[h!]
\smallskip
\begin{centering}
\includegraphics[scale=0.65]{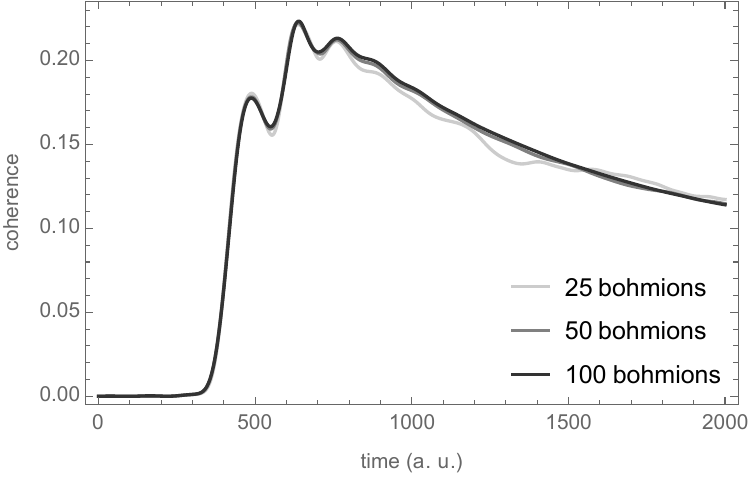}\includegraphics[scale=0.65]{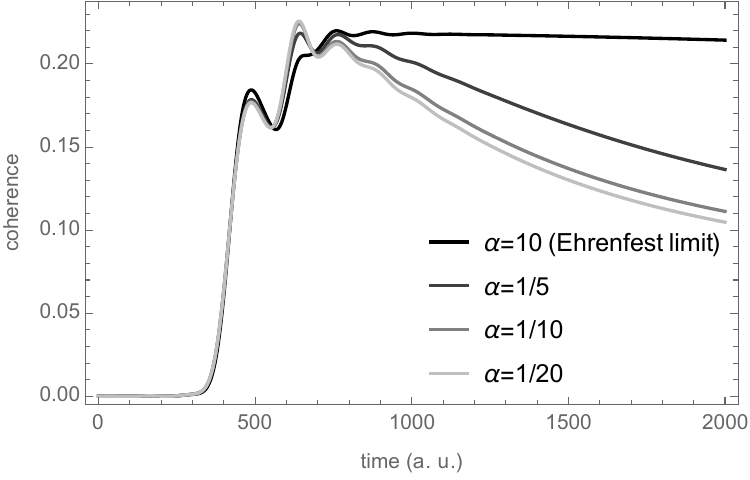}
\par\end{centering}
\caption{Left: coherence plots for varying $N$ and fixed $\alpha=1/10$,
showing good convergence for as few as $N=50$ bohmions. Right: converged (large $N)$ coherence plots for varying $\alpha$,
incorporating the large $\alpha$ Ehrenfest limit (classical trajectories
for mean-field dynamics) and showing convergence to the small $\alpha$
Bohmian limit (Bohmian trajectories for exact dynamics).}
\label{dependence}
\end{figure}
The
largest $\alpha=10$ corresponds to Ehrenfest dynamics, while, at
the other end, the small $\alpha\sim1/20$ trajectories accurately reproduce the full quantum result. We see that bohmions capture the qualitative
behaviour of the full quantum result even for $\alpha=1/5$, with
electronic decoherence effects (missed in Ehrenfest dynamics) visible
for $t>1000$. In general, of course, the appropriate choice of $\alpha$
and $N$ depends on the situation. In the next section, we consider a more challenging physical scenario.
}

\subsection{Tully III (extended coupling region with reflection)}
\begin{figure}[h]
\smallskip
\begin{centering}
\includegraphics[width=0.5\paperwidth]{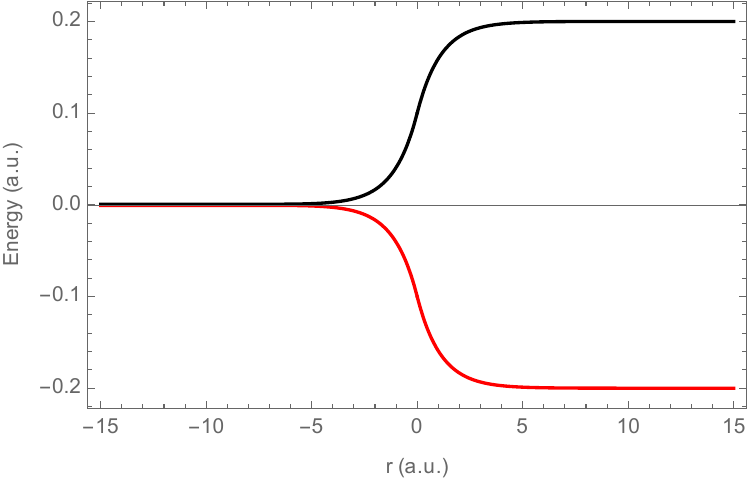}
\par\end{centering}
\caption{Tully III BO energy surfaces.\label{fig:TullyIIIBO}}
\end{figure}

Tully III is defined by the electronic matrix elements 
\begin{align}
H_{11}\left(r\right)&=-H_{22}\left(r\right)=a,
\\
H_{12}\left(r\right)&=b\left[2-\exp\left(-cr\right)\right],\quad r>0,
\\
H_{12}\left(r\right)&=b\exp\left(cr\right),\quad r<0,
\\
H_{12}\left(r\right)&=H_{21}\left(r\right)
\end{align}
with $a=0.0006,b=0.1,c=0.9$. The BO energy surfaces are illustrated
in Figure \ref{fig:TullyIIIBO}. 
The nuclear wavepacket first encounters an extended
coupling region ($r<0$), where the two BO energy surfaces are very close together, before the BO surfaces move apart. The wavepacket branches at this point, and (depending on the initial momentum) the part on the upper BO surface can be reflected if it doesn't have sufficient energy to climb the potential barrier. This reflected wavepacket then encounters the extended coupling region for a second time. It was suggested in \cite{Curchod} that these dynamics, involving a reflection process which leads to a second passage through a region of nonadiabatic coupling, are paralleled to some extent in the nonradiative deactivation of fulvene.

\begin{figure}[h]
\begin{centering}
\includegraphics[width=0.35\paperwidth]{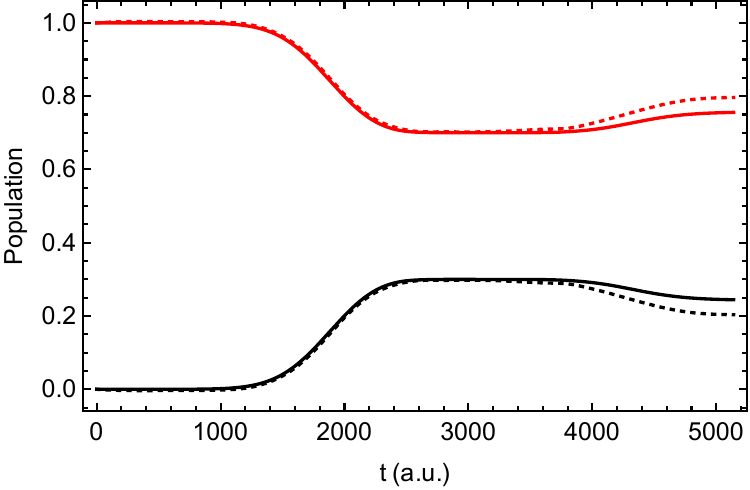}\includegraphics[width=0.35\paperwidth]{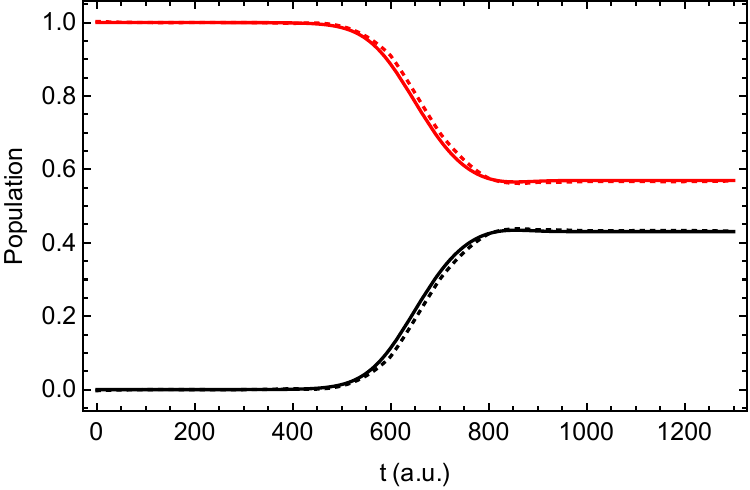}
\includegraphics[width=0.35\paperwidth]{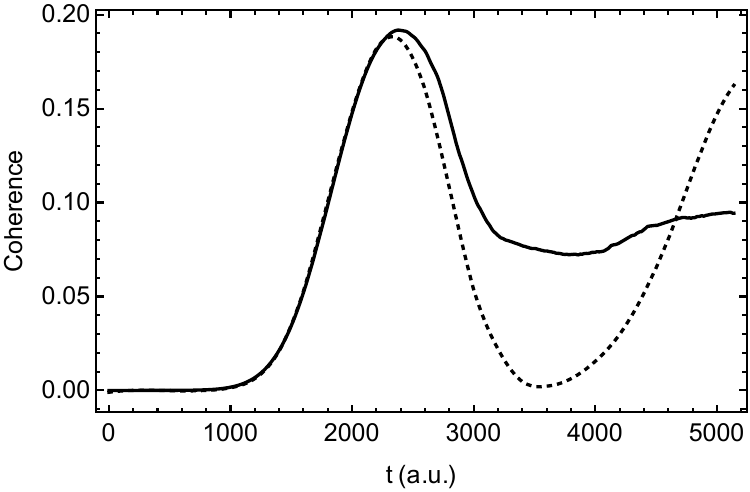}\includegraphics[width=0.35\paperwidth]{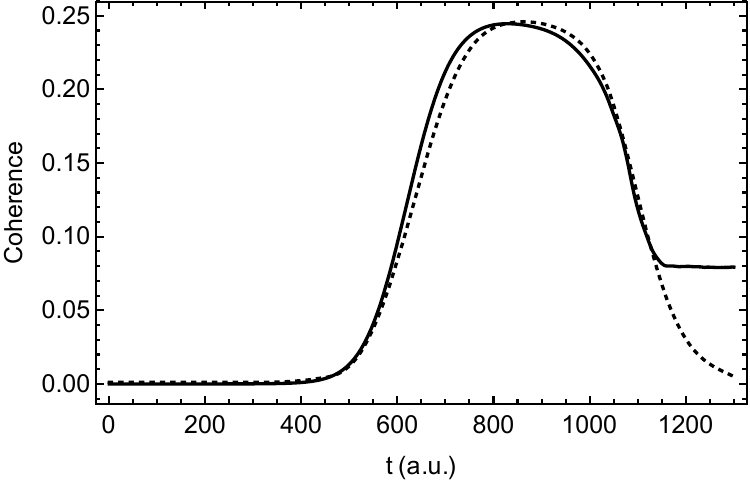}
\par\end{centering}
\caption{BO populations and coherence for Tully III. Left: $k=10$, ${N}=4000$,
${\alpha}=1/20$. Right: $k=30$, ${N}=2000$, ${\alpha}=1/30$.  The dotted lines are  the exact results for comparison. \label{fig:TullyIIIResults}}
\end{figure}
\paragraph{Decoherence.}
We {show simulations for nuclear wavepackets with} initial
momenta $k=10$ and $k=30$. {The centre of the initial wavepacket in \eqref{InMolWF} is set to $r_{0}=-15$ a.u. in both cases.} In Figure \ref{fig:TullyIIIResults} we
plot the BO populations and coherence measure in both cases. The bohmion simulations perform very well for the higher momentum
case $k=30$, again capturing the decoherence with {good accuracy}; see Figure \ref{fig:Grossplots} in Appendix \ref{app:comp}. 
The lower momentum
simulation involves more challenging dynamics, with significant wavepacket
splitting and reflection. The results shown in Figure \ref{fig:TullyIIIResults}, computed
with a regularization lengthscale of ${\alpha}={1}/{20}$ a.u., successfully
capture the correct qualitative behaviour of the coherence measure
throughout, although losing some accuracy at later times $>3000$
a.u. when the reflected wavepacket re-enters the extended coupling
region. {This loss in accuracy may be due to interference effects, which have been known to pose certain limitations to trajectory-based models \cite{ZhMa03}. As we shall see, a similar behaviour also occurs in the case of study treated in the next section.}

\paragraph{Accuracy.}
More accurate results can in principle be obtained by choosing a smaller
regularization lengthscale ${\alpha}$, as discussed earlier. However, {as mentioned earlier,} decreasing the regularization lengthscale ${\alpha}$ also comes at the cost of increasing
the number of bohmions and requiring a smaller timestep. In practice we find that the latter is the principle limitation, because our numerical method (based on a fixed timestep Runge-Kutta scheme) eventually loses stability for much smaller ${\alpha}$. It would be worth investigating whether this situation could be improved by using an adaptive scheme.

\subsection{Double Arch model\label{sec:DAmodel}}
\begin{figure}[h!]
\smallskip
\begin{centering}
\includegraphics[width=0.5\paperwidth]{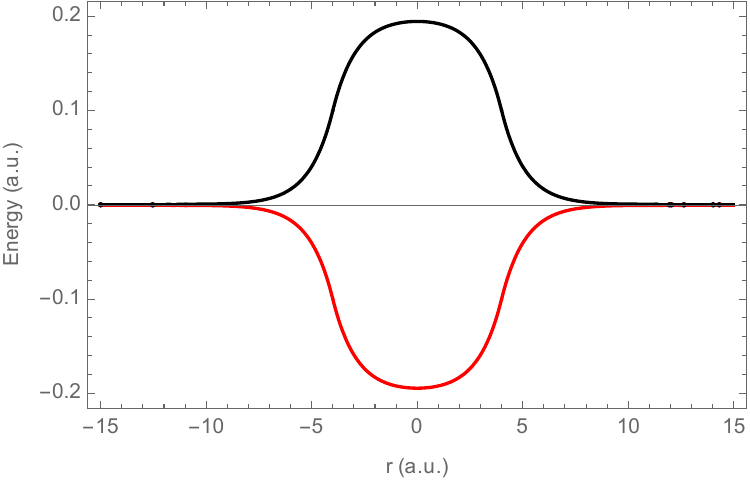}
\par\end{centering}
\caption{Double Arch model BO energy surfaces.\label{fig:DABO}}
\end{figure}

The Double Arch model is defined by the electronic matrix elements 
\begin{align}
H_{11}\left(r\right)&=-H_{22}\left(r\right)=a,
\\
H_{12}\left(r\right)&=-b\exp\left(c\left(r-d\right)\right)+b\exp\left(c\left(r+d\right)\right),\quad r<-d,
\\
H_{12}\left(r\right)&=b\exp\left(-c\left(r-d\right)\right)-b\exp\left(-c\left(r+d\right)\right),\quad r>d,
\\
H_{12}\left(r\right)&=2b-b\exp\left(c\left(r-d\right)\right)-b\exp\left(-c\left(r+d\right)\right),\quad -d<r<d,
\\
H_{21}\left(r\right)&=H_{12}\left(r\right)
\end{align}
with $a=0.0006,b=0.1,c=0.9,d=4$. The BO energy surfaces are illustrated
in Figure \ref{fig:DABO}. 
In this model, the lower and upper BO surfaces are initially close but move apart at $r\approx-5$ at which point we expect population transfer into the upper BO state and the wavepacket to split into two. 
\begin{figure}[h!]
\vspace{.5cm}
\begin{centering}
\includegraphics[width=0.35\paperwidth]{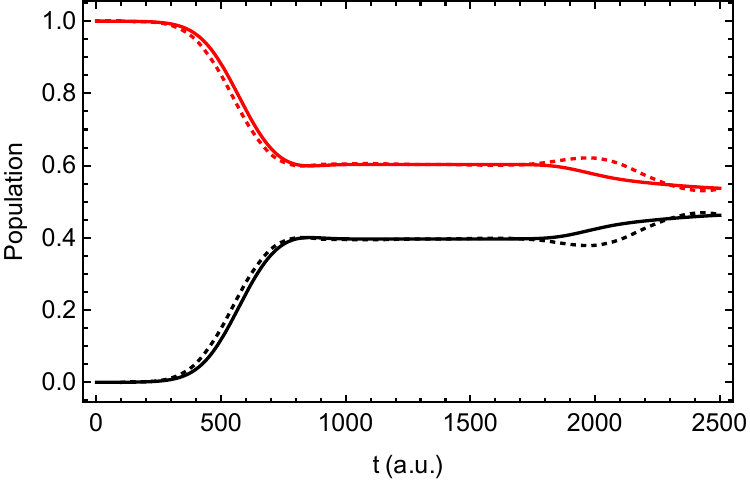}\includegraphics[width=0.35\paperwidth]{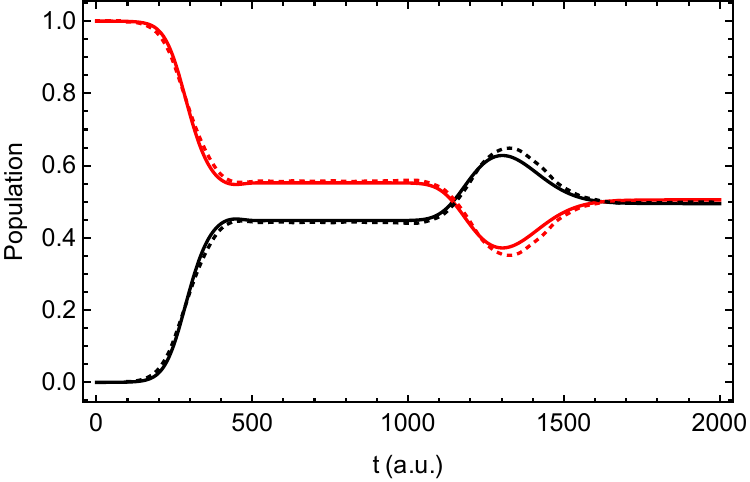}
\includegraphics[width=0.35\paperwidth]{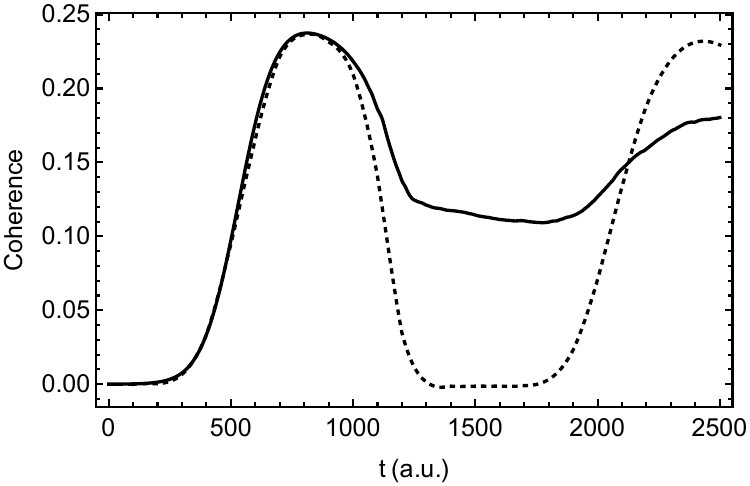}\includegraphics[width=0.35\paperwidth]{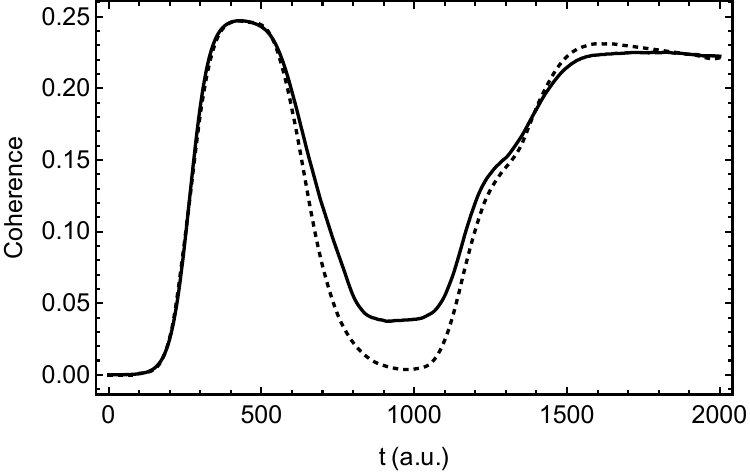}
\par\end{centering}
\caption{BO populations and coherence for Double Arch model. Left: $k=20$,
${N}=2000$, ${\alpha}=1/30$. Right: $k=40$, ${N}=2000$, ${\alpha}=1/30$.  The dotted lines are  the exact results for comparison.\label{fig:DAResults}}
\end{figure}
The wavepacket on the upper surface moves slower than the wavepacket on the lower surface, leading to spatial separation and significant decoherence as the wavepackets lose memory of each other.
 The two BO surfaces then come back together at $r\approx5$, causing further nonadiabatic transitions, at which point the wavepackets are recombined and interfere.
 
\paragraph{Decoherence.}
We run two simulations of nuclear wavepackets with given initial
momenta $k=20$ and $k=40$, respectively, {while we set  
 $r_{0}=-15$ a.u. in \eqref{InMolWF}}. In Figure \ref{fig:DAResults} we
plot the BO populations and coherence measure in both cases. Upon comparing again with Figure \ref{fig:Grossplots}, we see that the quality of agreement is similar to Tully III: the correct qualitative
behaviour is seen throughout the simulations, though with a loss of
accuracy, particularly for the lower momentum case ($k=20$) at later
times ($t>1000$). 

\paragraph{Dependence on $\boldsymbol\alpha$.}
{At this point, we present once again a more detailed discussion by emphasizing  specific dynamical features  obtained for the double-arch model, in this case with $k_0=40$. Here, we will compare our findings only with the results from the exact theory. \
We begin by analysing the $\alpha$-dependence of the coherence and population dynamics. Some illustrative plots are given in Figure \ref{archdependence}. In each of these, the exact quantum result is indicated by a dotted line. On looking at the decoherence dynamics, we see from the exact result that the coherence drops to virtually zero at $t\approx750$ as the wavepacket splits into two divergent wavepackets, one on each BO surface. Then, after $t\approx1000$, the wavepackets encounter the region where the BO surfaces come back together and the coherence measure goes back up due to further nonadiabatic transitions. 
\begin{figure}[h!]'
\smallskip
\begin{centering}
\includegraphics[scale=0.47]{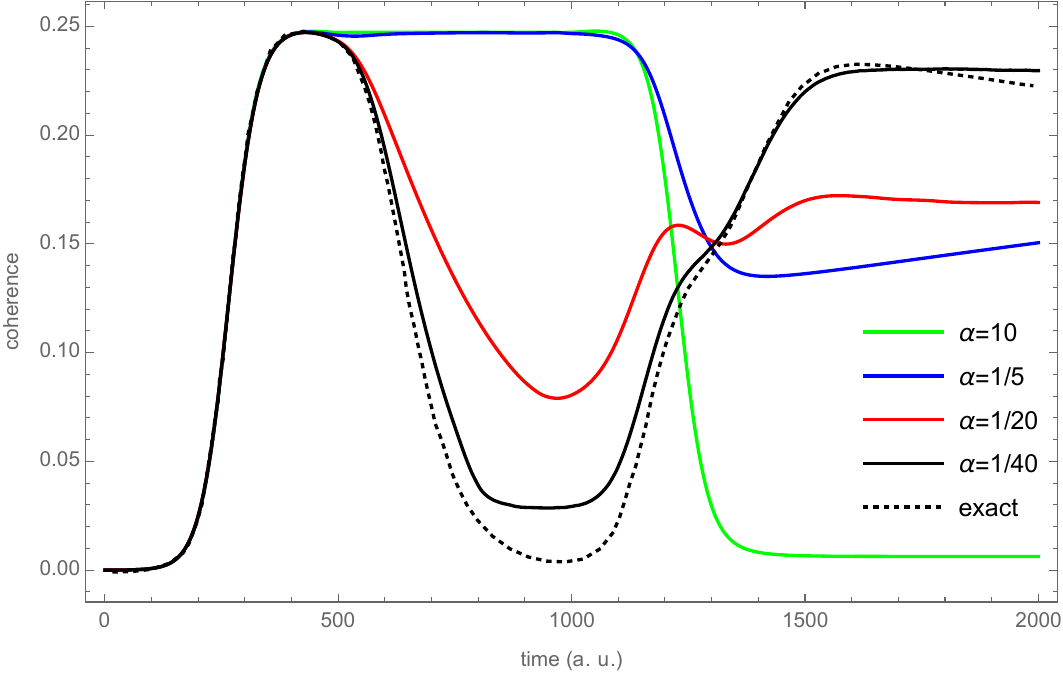}\includegraphics[scale=0.47]{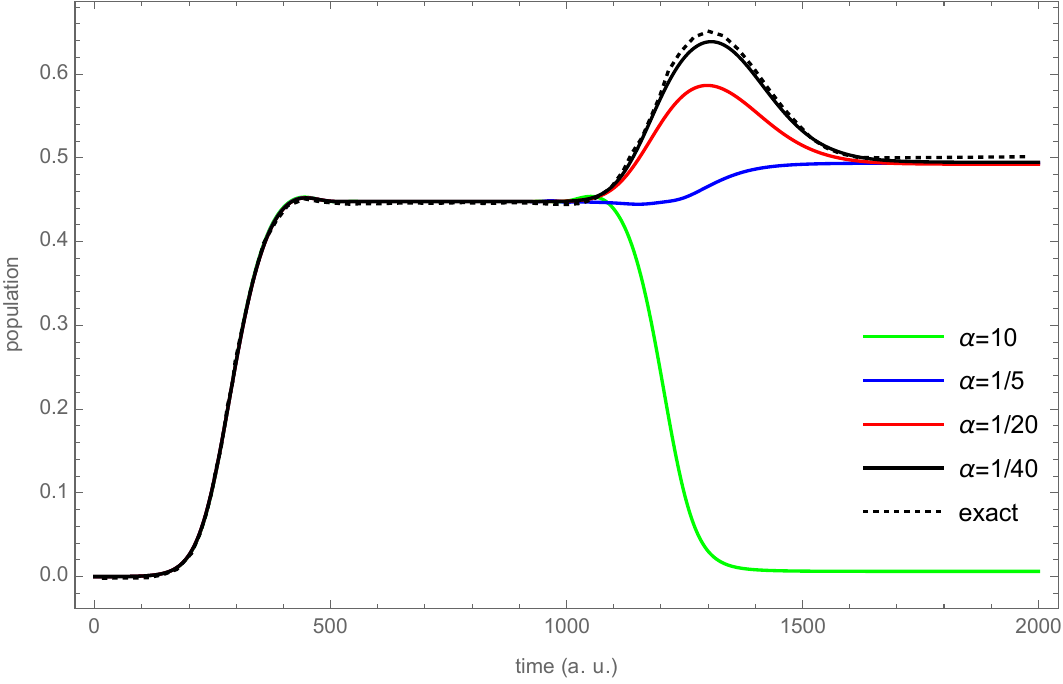}
\par\end{centering}
\caption{$\alpha$-dependence of bohmion dynamics for double arch model ($N=8000$). Left: coherence. Right: population.}
\label{archdependence}
\end{figure}

\paragraph{Comparison to Ehrenfest.}
We see that in the Ehrenfest limit ($\alpha=10$) and even for $\alpha=1/5$ this behaviour is missed by the bohmion dynamics. {Namely,} the wavepacket {remains intact} and no electronic decoherence is observed following the separation of the BO surfaces. {However,} the bohmion dynamics do capture the correct qualitative behaviour for $\alpha=1/20$ and {the bohmion results} are reasonably close to the exact result for $\alpha=1/40$. For this final value of $\alpha$, the splitting of the wavepacket is captured very effectively as illustrated in Figure \ref{split}. To summarise, we find that {to achieve the correct qualitative behaviour the bohmion method requires smaller $\alpha$ in wavepacket splitting} than in the dual avoided crossing model. 
\begin{figure}[!h]
\bigskip
\begin{centering}
\includegraphics[scale=0.47]{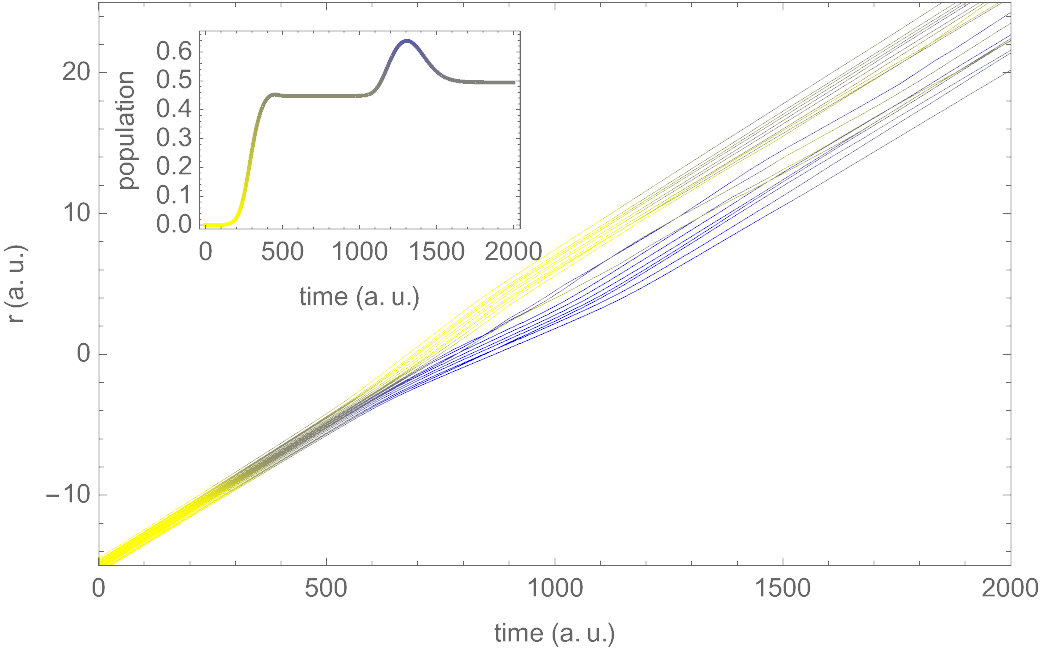}\includegraphics[scale=0.47]{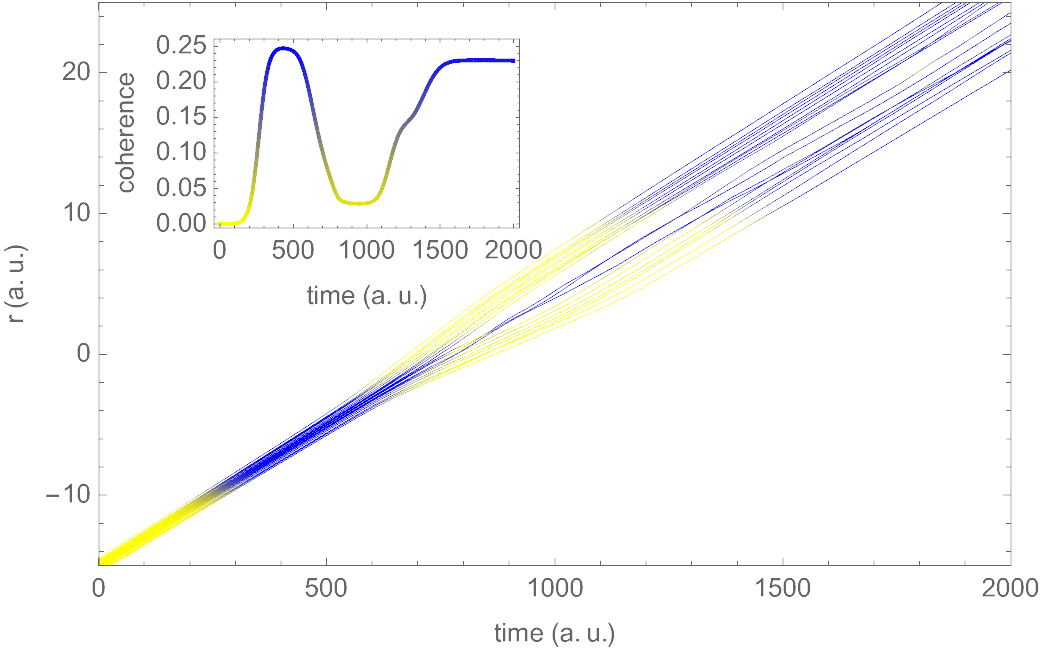}
\par\end{centering}
\caption{$\alpha=1/40,N=8000$ double arch model simulation. Left: population dynamics (yellow is lower BO state, blue is upper BO state). Right: coherence dynamics (appearance of yellow indicates decoherence). }
\label{split}
\end{figure}

\paragraph{Convergence.}
In the double arch model we also found in our simulation that a rather large number of bohmions ($N$) were required in order for convergence of the result. We found that $N<100$ was sufficient for good convergence} in the cases of the single and dual avoided crossing. However, for Tully III and the double arch model we found that thousands of bohmions were needed in some cases to achieve comparable convergence of the coherence measure. We found the convergence particularly slow for $t>1000$, indicating that quantum interference effects (which are relevant to the dynamics of the recombining wavepackets) may play an important role {in the convergence of the coherence measure}. This seems to be consistent with findings from numerical implementations of Bohmian trajectories, as well \cite{ZhMa03}, suggesting that quantum interference can lead to highly localized features in the probability density.  This, in turn, would require higher spatial resolution (i.e.  additional bohmions) to capture the physics. 

\section{Conclusions\label{sec:conclusions}}

In this work, we have applied the recently developed bohmion method to the celebrated Tully models of nonadiabatic molecular dynamics. Unlike other nonadiabatic methods based on hydrodynamic quantum trajectories, the bohmion method retains the fundamental conservation laws arising from its variational structure via Noether's theorem. We have compared the present scheme with other approximate approaches, including Ehrenfest, TSH and CT-MQC, as well as the exact quantum mechanical results. In the case of the Tully models, we were able to assess the extent to which these methods can accurately capture essential features of nonadiabatic dynamics such as population transfer and electronic decoherence. 

Our simulations have demonstrated that the bohmion method {can accurately capture electronic decoherence}.  {This is not unexpected, because the} bohmion method, being based on the exact factorization of the molecular wavefunction, retains correlations between the electrons and nuclei which are crucial to the decoherence dynamics through the inclusion of the (regularized) quantum potential, a non-local interaction potential which depends on the positions of all the bohmions.
The {bohmion} method performed best on Tully I and Tully II, with a loss of accuracy in the case of Tully III which involves wavepacket reflection at low wavepacket momenta. 
{To achieve sufficient accuracy, we employed a few thousand bohmions in our simulations, although only} a few hundred bohmions were sufficient to account for effects such as wavepacket reflection (in Tully III) even if with some loss of accuracy {in decoherence}.

Although the present results are encouraging, the bohmion method has by no means captured all the possible effects of quantum hydrodynamics. In particular, the interference patterns arising from highly irregular profiles of the quantum potential have been missed in  the present treatment, and we expect that capturing them will require further developments of the approach. 
{So far, we have also looked at the behaviour of bohmions (without electronic degrees of freedom) in a quartic well, and similarly found that {additional} bohmions were required to capture the relevant behaviour of local observables after one or two periods, at which point quantum interference effects become important.}
As interference patterns involve zeroes of the nuclear density, expansions of the type \eqref{BohmAns1} may not be appropriate in those cases, since the weights possess trivial dynamics. Improved alternative approaches may require retaining some phase information, perhaps by transporting both the phase and the amplitude of the nuclear wavefunction, or perhaps by transporting its Wigner transform, as opposed to the nuclear density alone. These open problems are beyond the  scope of the present work {and will} be investigated elsewhere.

\paragraph{Acknowledgements.}  
We would like to thank our friends and colleagues who have generously offered their time, thoughts and encouragement in the course of this work during the time of COVID-19. 
We are particularly grateful to Federica Agostini, Denys Bondar, Basile Curchod, Michael Foskett, and Dmitry Shalashilin for thoughtful suggestions and discussions.
The work of DDH and JIR was partially supported by the EPSRC grant CHAMPS EP/P021123/1. CT was partially supported by the Royal Society grant IES\textbackslash R3\textbackslash 203005. This work was undertaken on ARC4, part of the High Performance Computing facilities at the University of Leeds, UK.

\appendix

\bigskip

\section{Comparison with other methods\label{app:comp}}
\begin{figure}[!h]
\smallskip
\begin{centering}
\includegraphics[width=0.4\paperwidth]{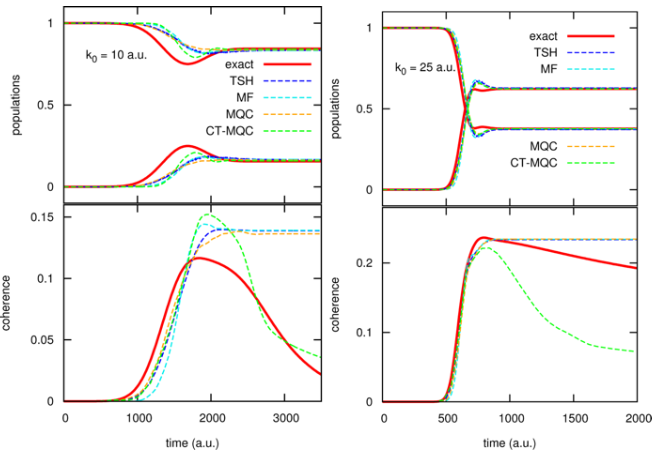}\includegraphics[width=0.4\paperwidth]{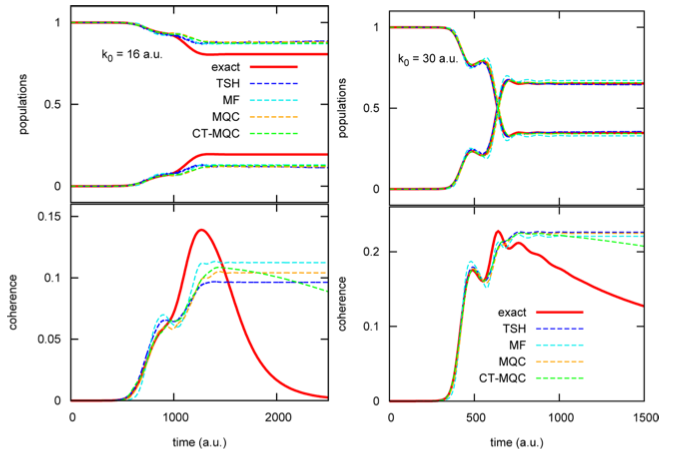}
\par\end{centering}
\begin{centering}
\includegraphics[width=0.4\paperwidth]{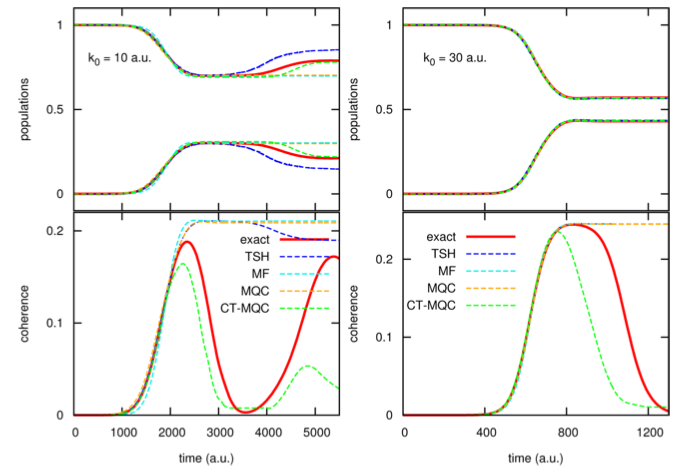}\includegraphics[width=0.4\paperwidth]{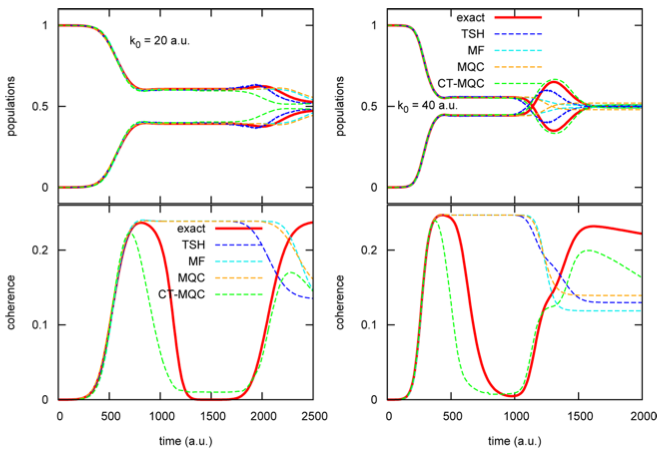}
\par\end{centering}
\caption{Results for (left to right; top to bottom) model (a), (b), (c) and
(d). Reprinted (adapted) with permission from F. Agostini, S.~K. Min, A. Abedi, E.~K.~U. Gross,  J. Chem. Theory Comput. 12 (2016), n. 5, 2127-2143. 
Copyright 2016 American Chemical Society.\label{fig:Grossplots}}
\bigskip
\end{figure}

Here, we display the results obtained on the Tully models by using other methods; see Figure \ref{fig:Grossplots}. In particular, we consider the Ehrenfest mean-field method (MF), the trajectory surface-hopping (TSH), and two variants of the  mixed quantum-classical method (MQC and CT-MQC). These plots appeared in  reference \cite{CoupledvsIndependent} and are presented here for comparison.

\end{document}